\def\tr{{\rm tr}\,}
\def\Tr{{\rm Tr}\,}

\def\sgn{{\rm sgn\,}}
\def\b{\bibitem}
\def\be{\begin{equation}}
\def\ee{\end{equation}}
\def\bea{\begin{eqnarray}}
\def\eea{\end{eqnarray}}
\def\bml{\begin{mathletters}}
\def\eml{\end{mathletters}}
\documentstyle[aps,prb,eqsecnum,psfig,epsf,floats]{revtex}
\draft
\begin{document}
\def\SNG{{\em Physical Review Style and Notation Guide}}
\def\LUG {{\em \LaTeX{} User's Guide \& Reference Manual}}
\def\btt#1{{\tt$\backslash$\string#1}}%
\def\REVTeX{REV\TeX}
\def\AmS{{\protect\the\textfont2
        A\kern-.1667em\lower.5ex\hbox{M}\kern-.125emS}}
\def\AmSLaTeX{\AmS-\LaTeX}
\def\BibTeX{\rm B{\sc ib}\TeX}
\twocolumn[\hsize\textwidth\columnwidth\hsize\csname@twocolumnfalse%
\endcsname
\title{Coexistence of ferromagnetism and superconductivity\\
       \small{$[$ Phys. Rev. B. {\bf 67}, 024515 (2003) $]$}
}
\author{T.R.Kirkpatrick}
\address{Institute for Physical Science and Technology, and Department of
         Physics\\
         University of Maryland,
         College Park, MD 20742}
\author{D.Belitz}
\address{Department of Physics and Materials Science Institute\\
         University of Oregon,
         Eugene, OR 97403}
\date{\today}
\maketitle

\begin{abstract}
A comprehensive theory is developed that describes the coexistence of
p-wave, spin-triplet superconductivity and itinerant ferromagnetism.
It is shown how to use field-theoretic techniques to derive both
conventional strong-coupling theory, and  analogous gap equations
for superconductivity induced by magnetic fluctuations. It is then
shown and discussed in detail that the magnetic fluctuations are
generically stronger on the ferromagnetic side of the magnetic
phase boundary, which substantially enhances the superconducting 
critical temperature in the ferromagnetic phase over that in the
paramagnetic one. The resulting phase diagram is compared with
the experimental observations in UGe$_2$ and ZrZn$_2$.
\end{abstract}
\pacs{PACS numbers: 74.20.Mn; 74.20.Dw; 74.62.Fj; 74.20.-z}
]
\section{Introduction}
\label{sec:I}

It is well known that, in principle, the exchange of magnetic fluctuations 
between electrons can induce superconductivity in both the paramagnetic and 
ferromagnetic phases of metals.\cite{AndersonBrinkman} In general, 
for superconductivity to occur 
one needs both some sort of attraction between quasi-particles, which can
be provided by magnetic flucutations, and low temperatures. Since magnetic 
fluctuations become large
near continuous magnetic phase transitions, ideal candidates for this
phenomenon would seem to be itinerant ferromagnets with a low Curie
temperature. In contrast to the much more common phonon-exchange case, which
usually leads to spin-singlet, s-wave superconductivity, the magnetically
mediated pairing is believed to usually be strongest in the spin-triplet,
p-wave channel. This type of superconductivity is very sensitive to
nonmagnetic disorder, so that very clean samples are 
also required. The
combined requirements of low temperatures, high purity, and vicinity to a
continuous ferromagnetic transition severely restricts the number of
materials where ferromagnetically induced superconductivity might be
observed. Furthermore, recent theoretical
work has shown that at low temperature the ferromagnetic transition in an
itinerant electron system should be generically of first 
order,\cite{us_1st_order,us_tbp} and there is some experimental evidence for 
this as well.\cite{experimental_1st_order} If this is the case, then there 
are no divergent magnetic fluctuations even at the transition point itself.

\begin{figure}[ht]
\epsfxsize=85mm
\centerline{\epsffile{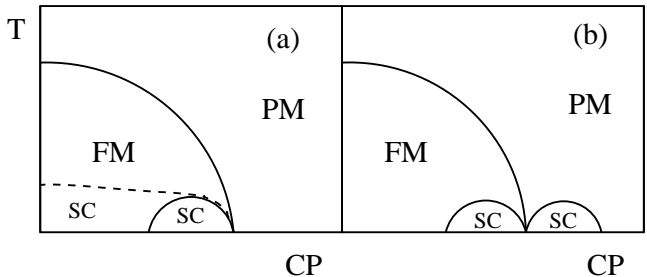}}
\vskip 5mm
\caption{Schematic phase diagram showing the paramagnetic (PM), ferromagnetic
 (FM), and superconducting phases (SC) in a temperature (T) - control parameter
 (CP) plane. (a) shows the qualitative PM-FM phase boundary (solid line)
 as well as the qualitative superconducting phase boundary line as observed in 
 UGe$_2$, Ref.\ \protect\onlinecite{Saxena_etal} (solid line) and
 ZrZn$_2$, Ref.\ \protect\onlinecite{Pfleiderer_etal} (dashed line),
 respectively, and explained by the theory
 presented here. In Ref.\ \protect\onlinecite{Saxena_etal}, hydrostatic 
 pressure serves as CP. (b) shows the qualitative prediction of paramagnon
 theory, Ref.\ \protect\onlinecite{FayAppel}.}
\vskip 0mm
\label{fig:1}
\end{figure}

Very recently, however, coexistence of ferromagnetism and 
superconductivity has indeed been observed, in 
UGe$_2$.\cite{Saxena_etal} The f-electron structure of the uranium
notwithstanding, this material is more similar to d-band metals than to heavy 
fermion systems, and ferromagnetism and superconductivity are believed to be
caused by itinerant electrons in the same band. The persistence of 
ferromagnetic 
order within the superconducting phase has been ascertained by neutron 
scattering. Since superconductivity in the presence of ferromagnetism is
likely to be of spin-triplet type, magnetic-fluctuation induced pairing is a
possible mechanism. However, two aspects of the experiments are not readily
reconcilable with previous theories of superconductivity in ferromagnets.
First, the magnetic
transition is observed to be of first order, so that there are no divergent
fluctuations as the transition is approached. Second, the superconducting
state is found only on the ferromagnetic side of the phase boundary. A
schematic phase diagram for this situation is shown in Fig.\ \ref{fig:1}(a).
The theoretical prediction, on the other hand, has been a phase diagram
like the one shown in Fig.\ \ref{fig:1}(b), with superconductivity existing, 
more or less symmetrically, on both sides of the magnetic phase 
boundary.\cite{FayAppel,anomaly_footnote}

Since the initial experimental observation of superconductivity in UGe$_2$
there has been considerable new experimental and theoretical work. 
Experimentally, the qualitatively same phenomenon has been observed in 
ZrZn$_2$,\cite{Pfleiderer_etal} and in URhGe.\cite{Aoki_etal} In
ZrZn$_2$, the paramagnetic-to-ferromagnetic phase transition appears to 
be either continuous or only weakly first order; in URhGe the nature
of the magnetic transition at very low temperatures is not known. In
both of these materials, superconductivity is
observed far inside the ferromagnetic phase. This is in contrast to UGe$_2$,
where it is observed only close to the magnetic phase boundary, see
Fig.\ \ref{fig:1}(a).
Theoretically, Shick and Pickett\cite{ShickPicket} have included 
correlation effects in a local density approximation
approach and have concluded that magnetically mediated triplet
superconductivity can occur in UGe$_2$. In contrast, Shimahara and 
Kohmoto\cite{ShimaharaKohmoto} have argued that triplet superconductivity 
can be induced by screened phonon interactions in ferromagnetic compounds 
such as UGe$_2$. Santi et al.\cite{Santi_et_al} have
performed ab initio calculations for ZrZn$_2$, concluding that
magnetically mediated triplet superconductivity can occur, with a critical
superconductivity temperature as high as 1 K. Watanabe and 
Miyake\cite{WatanabeMiyake} have
argued that coupled charge density waves and spin density waves can be used
to understand the superconductivity in UGe$_2$, and a 
Larkin-Ovchinnikov-Fulde-Ferrell state is also a 
possibility.\cite{Blagoev_et_al}
Machida and Ohmi,\cite{MachidaOhmi} and more recently Samokhin and 
Walker,\cite{SamokhinWalker} have investigated possible order parameters
for magnetically induced superconductivity, and a $d$-electron exchange
mechanism has been proposed.\cite{SpalekWrobel} Finally, the experiments
have prompted investigations of the detailed role of the normal self
energy in the Eliashberg equations for magnetically induced 
superconductivity.\cite{Wang_et_al,RoussevMillis,finite_Tc_footnote}

In a previous short paper,\cite{us_letter} the current authors and collaborators
have shown that the critical
temperature for spin-triplet, p-wave superconductivity mediated by spin 
fluctuations is generically much higher in a Heisenberg ferromagnetic
phase than in a paramagnetic one, due to the coupling of the magnons to the
longitudinal magnetic susceptibility, and that this qualitatively explains
the observed phase diagram. In the current paper we explain and expand on 
this work. There are two important aspects. The first, formal one, is to
show how to use field-theoretic techniques, rather than many-body
perturbation theory, to derive Eliashberg-like 
equations for magnetic-fluctuation induced superconductivity. 
For completeness, and to motivate the more complicated
case of magnetic-fluctuation induced superconductivity, we also show 
how these ideas can be used to derive conventional Eliashberg theory.
The second one is a detailed explanation of our previous observation
that magnetic fluctuations are generically stronger on the ferromagnetic 
side of the phase boundary, which in turn is shown to
lead to a much higher critical temperature for superconductivity in the 
ferromagnetic phase. This result is consistent with the
observed phase diagrams in UGe$_2$ and ZrZn$_2$, schematically shown in
Fig.\ \ref{fig:1}(b). The formalism developed here will also lay the ground
for further investigations of the superconductor-ferromagnet phase diagram. 
In particular, we will argue that as a function of temperature and
magnetization, one expects phase transitions to other, more exotic,
superconducting states inside the superconducting phase.

The plan of this paper is as follows. In Section\ \ref{sec:II} we derive
an effective action or field theory for superconductivity due to magnetic 
fluctuations. The field theory is made explicit by keeping magnetic 
fluctuations to one-loop order. The resulting superconducting field theory 
is then solved in a saddle-point approximation in Section\ \ref{sec:III}. 
The resulting gap equations are consistent with earlier 
results,\cite{FayAppel}
except that the interaction is an exact longitudinal magnetic susceptibility 
rather than an approximate one. The field theoretic derivation also has
structural advantages that we plan to exploit in the future, as is discussed
in Section\ \ref{sec:V}. In the last part of Section \ref{sec:III}, explicit 
expressions for the magnetic susceptibility in either phase are given. 
Emphasis is on the difference between magnetic fluctuations in the two 
phases. In Section\ \ref{sec:IV} the linearized gap equations derived in 
Section\ \ref{sec:III} are solved using a McMillan-type approximation. 
Phase diagrams are derived showing the paramagnetic, ferromagnetic and 
superconducting phases. It is concluded that the critical temperature for 
superconductivity is much higher in the ferromagnetic phase than in the 
paramagnetic phase. The paper is concluded in Section\ \ref{sec:V} with 
a discussion of our results, of possible future applications of the formalism 
developed here, and of some open questions. The derivation of conventional
Eliashberg theory by field-theoretic methods is relegated to an appendix.

\section{Coupled Field Theory for Normal State, Superconducting, and 
         Magnetic Fluctuations}
\label{sec:II}

\subsection{The model}
\label{subsec:II.A}

Our starting point is an interacting electron gas model. In terms of
Grassmann variables, the partition function is
\be
Z = \int D[{\bar\psi},\psi]\ e^{S[{\bar\psi},\psi]}\quad.  
\label{eq:2.1}
\ee
Here the functional integration is with respect to Grassmann-valued 
fields $\bar\psi$ and $\psi$, and the action $S$ reads
\bml
\label{eqs:2.2}
\be
S = -\int dx \sum_{\sigma}{\bar\psi}_{\sigma}(x)\,\partial_{\tau}\,
          \psi_{\sigma}(x) + S_{0} + S_{\rm int}\quad.  
\label{eq:2.2a}
\ee
We use a $(d+1)$-vector notation, with $x \equiv ({\bf x},\tau)$, and 
$\int dx \equiv \int_{V}d{\bf x}\int_{0}^{\beta}d\tau$. ${\bf x}$ denotes
position, $\tau$ imaginary time, $V\rightarrow\infty$ the system
volume, and $\beta =1/T$ the inverse temperature. $\sigma$ is a spin label, 
and we use units such that $\hbar = k_{\rm B} = 1$.
$S_{0}$ describes free electrons with chemical potential $\mu$,
\be
S_{0} = \int dx\sum_{\sigma} {\bar\psi}_{\sigma}(x)\left(
          \frac{\nabla^2}{2m_{\rm e}} + \mu\right) \psi_{\sigma}(x)\quad,  
\label{eq:2.2b}
\ee
with $m_{\rm e}$ the electron mass. $S_{\rm int}$ denotes the two-body 
electron-electron interaction. The underlying Coulomb 
interaction becomes screened, and after renormalization from a 
microscopic to a mesoscopic scale, $S_{\rm int}$ separates into 
spin-singlet and spin-triplet particle-hole and particle-particle 
interactions, respectively.\cite{AGD} The interaction in the 
particle-particle channel is what causes superconductivity, and we
do {\em not} include it in our bare action. Rather, one of the main 
points of this paper is to see how this interaction is generated by magnetic 
fluctuations. The interactions in the particle-hole channel are
\be
S_{\rm int} = S_{\rm s}^{\rm p-h} + S_{\rm t}^{\rm p-h}\quad,  
\label{eq:2.2c}
\ee
with,
\bea
S_{\rm s}^{\rm p-h} = \frac{\Gamma_s}{2}\int dx\ n(x)\,n(x)\quad,  
\label{eq:(2.2d)}\\
S_{\rm t}^{\rm p-h} = \frac{\Gamma_t}{2}\int dx\ {\bf n}_{\rm s}(x)
                      \cdot {\bf n}_{\rm s}(x)\quad.
\label{eq:2.2e}
\eea
\eml%
Here $\Gamma_{\rm s,t}$ are the particle-hole spin-singlet and 
spin-triplet interaction amplitudes, respectively.
\bml
\label{eqs:2.3}
\be
n(x) = \sum_{\sigma}{\bar\psi}_{\sigma}(x)\,\psi_{\sigma}(x)\quad,  
\label{eq:2.3a}
\ee
is the electron number density field, and
\be
n_{\rm s}^{i}(x) = \sum_{a,b}{\bar\psi}_{a}(x)\sigma_{i}^{ab}\psi_{b}(x)
                    \quad,\quad (i=1,2,3) \quad,  
\label{eq:2.3b}
\ee
\eml%
are the components of the electron spin density field ${\bf n}_{\rm s}$.
$\sigma_{1,2,3}$ denote the Pauli matrices.

Below we will often use a Fourier representation of the Grassmann field,
\bml
\label{eqs:2.4}
\bea
\psi_{\sigma,n}({\bf k}) &=& \sqrt{\frac{T}{V}}\int dx\ 
   e^{i(\omega_n\tau - {\bf k}\cdot{\bf x})}\ \psi_{\sigma}(x)\quad,  
\label{eq:2.4a}\\
{\bar\psi}_{\sigma,n}({\bf k}) &=& \sqrt{\frac{T}{V}}\int dx\ 
   e^{-i(\omega_n\tau - i{\bf k}\cdot{\bf x})}\ {\bar\psi}_{\sigma}(x)\quad.
\label{eq:2.4b}
\eea
\eml%
Finally, to introduce the relevant order parameters, it will be convenient
to introduce generalized Nambu spinors,
\bml
\label{eqs:2.5}
\bea
\Psi(x) &=& \frac{1}{\sqrt{2}}\left(\begin{array}{c}
                                     \psi_{\uparrow}(x)\\
                                     \psi_{\downarrow}(x)\\
                                     {\bar\psi}_{\uparrow}(x)\\
                                     {\bar\psi}_{\downarrow}(x)
                                    \end{array}\right)\quad,
\label{eq:2.5a}\\
{\bar\Psi}(x) &=& \frac{1}{\sqrt{2}}\,\left({\bar\psi}_{\uparrow}(x),
     {\bar\psi}_{\downarrow}(x),\psi_{\uparrow}(x),\psi_{\downarrow}(x)
     \right)\quad.  
\label{eq:2.5b}
\eea
and corresponding composite variables as the tensor product of the spinors,
\be
\Phi_{ij}(x,y) = {\bar\Psi}_{i}(x)\,\Psi_{j}(y)\quad.
\label{eq:2.5c}
\ee
\eml%
It is also useful to define a four-vector of $4\times 4$ matrices
\be
(\gamma_0,{\vec\gamma}) = (\sigma_3\otimes\sigma_0,\sigma_3\otimes\sigma_1,
                      \sigma_0\otimes\sigma_2,\sigma_3\otimes\sigma_3)\quad,
\label{eq:2.6}
\ee 
with $\sigma_0$ the $2\times 2$ unit matrix.\cite{gamma_footnote} 
In terms of these spinors and the 
$\gamma$, the noninteracting part of the action reads
\bml
\label{eqs:2.7}
\be
S - S_{\rm int} = \int dx\ {\bar\Psi}(x)\,\left[-\partial_{\tau}
                  + \gamma_0\left(\nabla^2/2m_{\rm e} + \mu\right)\right]
                       \,\Psi(x)\ ,
\label{eq:2.7a}
\ee
and the number density and spin density fields that make up the
interacting part can be written
\bea
n(x)&=&{\bar\Psi}(x)\gamma_0\Psi(x)\quad,
\label{eq:2.7b}\\
n_{\rm s}^{i}(x)&=&{\bar\Psi}(x)\gamma_i\Psi(x)\quad,\quad (i=1,2,3)\quad,
\nonumber\\
\label{eq:2.7c}
\eea
\eml%
with the product between spinors defined as the matrix product.

\subsection{Superconducting and magnetic order parameter field theory}
\label{subsec:II.B}

\subsubsection{Intuitive argument for magnetic-fluctuation induced
superconductivity}
\label{subsubsec:II.B.1}

To motivate our technical considerations, we recapitulate a simple
physical argument given by De Gennes.\cite{DeGennes}
Let us suppose that we have a local
magnetization fluctuation, $\delta{\bf M}({\bf x})$, that couples to the
electronic spin density. Since the coupling of $\delta {\bf M}$ to 
${\bf n}_{\rm s}$ is analogous to that of a magnetic field,
the change in energy of the electron system is given by a Zeeman term,
\bml
\label{eqs:2.8}
\be
\delta H = -\Gamma\int d{\bf x}\ {\bf n}_{\rm s}({\bf x})\cdot 
               \delta{\bf M}({\bf x})\quad,  
\label{eq:2.8a}
\ee
with $\Gamma$ a coupling strength, and ${\bf n}_{\rm s}({\bf x})$ the
electronic spin density. To first order in $\Gamma$, the magnetization 
induced at point ${\bf x}$ by an electronic spin density at point ${\bf y}$ 
is given by
\be
\delta M_{i}({\bf x}) = \Gamma\int d{\bf y}\sum_j\chi_{ij}({\bf x}-{\bf y})\,
    n_{\rm s}^{j}({\bf y})\quad,  
\label{eq:2.8b}
\ee
\eml%
where $\chi_{ij}$ is the electronic magnetic susceptibility tensor. 
This implies
\be
\delta H = -\Gamma^2 \int d{\bf x}\,d{\bf y}\sum_{ij} n_{\rm s}^{i}({\bf x})\,
           \chi_{ij}({\bf x}-{\bf y})\,n_{\rm s}^{j}({\bf y})\quad.
\label{eq:2.9}
\ee
Equation (\ref{eq:2.9}) shows that there is an 
effective electronic spin-spin interaction potential tensor given by
\be
V_{ij}({\bf x} - {\bf y}) = -\Gamma^2\,\chi_{ij}({\bf x}-{\bf y})\quad.
\label{eq:2.10}
\ee

This result has two important features that motivate our formal considerations
below. First, we anticipate that the pairing potential responsible for
superconductivity will be proportional to the magnetic susceptibility.
Second, the interaction will be attractive if the spins form a relative
spin-triplet, and repulsive if they form a 
spin-singlet.\cite{singlet_triplet_footnote}

\subsubsection{Order parameter field theory}
\label{subsubsec:II.B.2}

The above considerations motivate transforming the fermionic field theory 
to an effective one in terms of an order parameter field for spin-triplet
superconductivity, and a magnetization field. Since we are considering
an electronic mechanism for superconductivity, a (relatively) high $T_{\rm c}$
can be achieved  
only if electron-electron interaction effects are in some sense strong. 
This implies that for a consistent theory we need to also worry about 
strong-coupling effects as described by normal-state self energies, i.e., 
an Eliashberg-type theory is needed. We will accomplish this by introducing 
an additional field related to normal-state effects.

Because we will eventually use a saddle-point, or free energy variational,
technique we will have to choose a superconducting order parameter as an input. 
In the ferromagnetic phase it seems likely that the triplet
superconducting phase with the lowest energy will be the one with the spins
aligned in the direction of the magnetization. In what follows we assume 
that this is the case. We plan to also investigate other superconducting 
order parameters in the future, see Section \ref{sec:V} below.

In Appendix \ref{app:A} we show how field-theoretic techniques can be used 
to derive standard strong coupling theory for conventional superconductors. 
Here we essentially repeat that procedure for the order parameter field
mentioned above, with one modification. Because the Cooper interaction in the 
spin-triplet channel must be generated by magnetic fluctuations, we can no
longer use a saddle-point approximation for all the degrees of freedom.
Further, if the magnetic transition is of first order, this
also manifests itself as a fluctuation effect in this description. 
Below we will see that it is sufficient to keep magnetic fluctuations to 
one-loop order in order to derive a theory for magnetic-fluctuation induced 
spin-triplet superconductivity. Since the magnetic transition 
is already understood, we will model its first order nature in a simple way.
Generally, our guiding philosophy will be to keep fluctuations to leading order
if they lead to a qualitatively new effect that is relevant for the 
superconducting transition temperature, but to neglect them otherwise.

The first formal step is to introduce the magnetization field, ${\bf M}$,
by performing a Hubbard-Stratonovich transformation
on the triplet interaction term,
Eq.\ (\ref{eq:2.2e}). The other order parameter fields are introduced via a 
delta-function constraint as in Appendix \ref{app:A}, and we integrate out
the fermions.\cite{procedure_footnote} 
This allows us to write the partition function, in analogy
to Eq.\ (\ref{eq:A.4}), as
\be
Z = \int D[M,{\cal G},\Lambda]\ e^{{\cal A}[M,{\cal G},\Lambda]}\quad,  
\label{eq:2.11}
\ee
with the action ${\cal A}$ given by
\bml
\label{eqs:2.12}
\bea
{\cal A}[M,{\cal G},\Lambda] &=& \frac{\Gamma_s}{2}\int dx\left(\tr\left[
   \gamma_0{\cal G}(x,x)\right]\right)^2 
            + \Tr(\Lambda{\cal G})
\nonumber\\
&&\hskip -25pt +\frac{1}{2}\Tr\ln ({\tilde G}_{0}^{-1} + \sqrt{2\Gamma_{\rm t}}
   {\vec\gamma}\cdot{\bf M} - \Lambda^{\rm T}) 
\nonumber\\
&& -\int dx\ {\bf M}(x)\cdot {\bf M}(x)\quad.
\label{eq:2.12a}
\eea
Here we have constrained the field $\Phi$, Eq.\ (\ref{eq:2.5c}), to a bosonic
field ${\cal G}$ by means of the Lagrange multiplier $\Lambda$, and the
superscript T denotes a transposition operation. $\Tr$ is an operator that
traces over all degrees of freedom, including the continuous space-time
index, while $\tr$ denotes a trace over discrete indices only.
$(\gamma_0,{\vec\gamma})$ 
is the matrix-valued four-vector defined in Eq.\ (\ref{eq:2.6}),
and we have defined the inverse free electron Green operator,
\be
{\tilde G}_{0}^{-1} = -\partial_{\tau} + \gamma_0\left({\nabla^2}/2m_{\rm e} 
              + \mu\right)\quad.  
\label{eq:2.12b}
\ee
\eml%

The Eqs.\ (\ref{eq:2.11}) - (\ref{eqs:2.12}) are formally exact. To proceed,
we make a number of simplifications and approximations. First, we will 
assume a magnetic state, on the ferromagnetic (FM) side of the magnetic 
phase boundary, with an average magnetization $m$ in the z-direction. 
We thus write
\bml
\label{eqs:2.13}
\be
{\bf M}(x) = m\sqrt{\Gamma_{\rm t}/2}\,{\hat{\bf z}} + \delta{\bf M}(x)\quad.  
\label{eq:2.13a}
\ee
Motivated by the considerations at the beginning of the current subsection,
we assume a non-unitary triplet superconducting order 
parameter $\langle{\bar\psi}_{\uparrow}({\bf x})\,
{\bar\psi}_{\uparrow}({\bf y})\rangle$.\cite{beta_phase_footnote}  
In terms of the matrix elements of ${\cal G}$, which are isomorphic to those
of $\Phi$, this order parameter and its hermitian adjoint are
\bea
2\langle{\cal G}_{13}(x,y)\rangle \equiv F^{+}(x-y)\neq 0\quad,
\nonumber\\
2\langle{\cal G}_{31}(x,y)\rangle \equiv F(x-y)\neq 0\quad.  
\label{eq:2.13b}
\eea
We also define anomalous self energies, or gap functions,
\bea
\langle\Lambda_{13}(x,y)\rangle \equiv \Delta^{+} (x-y)\neq 0\quad,
\nonumber\\
\langle\Lambda_{31}(x,y)\rangle \equiv \Delta (x-y)\neq 0\quad.
\label{eq:2.13c}
\eea
The normal state order parameters and self energies are defined analogously
to those in Appendix \ref{app:A}, except that isotropy in spin space is lost 
in the ferromagnetic state,
\bea
2\langle{\cal G}_{11}(x,y)\rangle = - 2\langle{\cal G}_{33}(y,x)\rangle
                                 \equiv G_{\uparrow}(x-y)\quad,
\nonumber\\
2\langle{\cal G}_{22}(x,y)\rangle = - 2\langle{\cal G}_{44}(y,x)\rangle
                                 \equiv G_{\downarrow}(x-y)\quad,
\label{eq:2.13d}
\eea
\bea
\langle\Lambda_{11}(x,y)\rangle = - \langle\Lambda_{33}(y,x)\rangle
                                \equiv \Sigma_{\uparrow}(x-y)\quad,
\nonumber\\
\langle\Lambda_{22}(x,y)\rangle = - \langle\Lambda_{44}(y,x)\rangle
                                \equiv \Sigma_{\downarrow}(x-y)\quad.  
\label{eq:2.13e}
\eea
\eml%
Note that we are assuming that the equilibrium state is spatially
homogeneous. For our present discussion, which will focus on the superconducting
phase transition, this assumption is legitimate. However, within the 
superconducting phase it cannot be correct
in general, as we discuss in Section \ref{sec:V}.

As already stressed, the crucial interactions leading to the superconducting
state and the most important normal-state self-energy effects are due to
magnetic fluctuations. To take these fluctuations into account, we insert
Eq. (\ref{eq:2.13a}) into Eq.\ (\ref{eq:2.12a}), expand to order
$(\delta{\bf M})^2$, and then integrate out the magnetic fluctuations. 
The partition function and action can then be written
\be
Z = \int D[{\cal G},\Lambda]\ e^{A[{\cal G},\Lambda]}\quad,  
\label{eq:2.14}
\ee
with
\bml
\label{eqs:2.15}
\bea
A[{\cal G},\Lambda] &=& \frac{\Gamma_s}{2}\int dx\left(\tr\left[
   \gamma_0\,{\cal G}(x,x)\right]\right)^2
            + \Tr(\Lambda{\cal G})
\nonumber\\
&& + \frac{1}{2}\,\Tr\ln {\tilde G}^{-1}_{\Lambda} 
    - \frac{V\Gamma_{\rm t}m^2}{2T} 
   - \frac{1}{2}\,\Tr\ln\chi^{-1}_{\Lambda}\quad.
\nonumber\\
\label{eq:2.15a}
\eea
Here we have defined a $4\times 4$ matrix Green operator,
\be
{\tilde G}^{-1}_{\Lambda} = {\tilde G}_{0}^{-1} + \Gamma_{\rm t}m\gamma_3 
     - \Lambda^{\rm T}\quad,  
\label{eq:2.15b}
\ee
and a normalized inverse magnetic susceptibility tensor,
\bea
(\chi_{\Lambda}^{-1})_{ij}(x-y)&=&\delta_{ij}\delta(x-y)
\nonumber\\
&&\hskip -45pt + \frac{\Gamma_{\rm t}}{2}\,
    \tr\left[{\tilde G}_{\Lambda}(y,x)\gamma_i\,{\tilde G}_{\Lambda}(x,y)
                                                      \gamma_j\right]\ ,\ 
(i,j = 1,2,3)\ ,
\nonumber\\
\label{eq:2.15c}
\eea
\eml%
Notice that ${\tilde G}$ and $\chi$ both depend on $\Lambda$, as we 
explicitly indicate with our notation.

The action as written in Eqs.\ (\ref{eqs:2.15}) does not yet allow for a
saddle-point solution that describes a superconducting state, i.e. one with
a broken $U(1)$ gauge symmetry that corresponds to particle number conservation.
This can be seen by taking the variation with
respect to, say, ${\cal G}_{13}$, which yields
$\Lambda_{31}=0$, which characterizes a non-superconducting state.
Indeed, such a solution is more than just a saddle-point solution, as can be
seen by formally integrating out ${\cal G}_{13}$. However, while always a
solution of the field theory, such a state is not necessarily stable. 
To allow for superconducting states with a spontaneously broken gauge symmetry,
we write $\Lambda$ as its average value plus a fluctuation,
\be
\Lambda = \langle\Lambda\rangle + \delta\Lambda\quad.  
\label{eq:2.16}
\ee
We then formally integrate over the $\delta\Lambda$ fluctuations to
obtain
\bml
\label{eqs:2.17}
\be
Z = \int D[{\cal G}]\ e^{A[{\cal G},\langle\Lambda\rangle]}\quad,  
\label{eq:2.17a}
\ee
with
\bea
A[{\cal G},\langle\Lambda\rangle] &=& \frac{\Gamma_s}{2}\int dx\left(\tr\left[
   \gamma_0{\cal G}(x,x)\right]\right)^2
            + \Tr(\langle\Lambda\rangle\,{\cal G})
\nonumber\\
&& +\frac{1}{2}\,\Tr\ln {\tilde G}^{-1}_{\langle\Lambda\rangle}
   -\frac{1}{2}\,\Tr\ln\chi^{-1}_{\langle\Lambda\rangle} 
    - \frac{V\Gamma_{\rm t}m^2}{2T}
\nonumber\\
&&\hskip -15pt +\ln\int D[\delta\Lambda]\exp\left\{\Tr({\cal G}\delta\Lambda)
   + \frac{1}{2}\,\Tr\ln\frac{{\tilde G}^{-1}_{\langle\Lambda\rangle 
                                                         + \delta\Lambda}}
        {{\tilde G}^{-1}_{\langle\Lambda\rangle}}
     \right.
\nonumber\\
&& \hskip 50pt\left. -\frac{1}{2}\,
    \Tr\ln\frac{\chi^{-1}_{\langle\Lambda\rangle + \delta\Lambda}}
        {\chi^{-1}_{\langle\Lambda\rangle}}\right\} \quad.
\label{eq:2.17b}
\eea
\eml%
Notice that $\langle\Lambda_{13}\rangle = \Delta^{+}$, which
couples to ${\cal G}_{31}$ in the $\Tr (\langle\Lambda\rangle\,{\cal G})$ 
term, acts like an external field that
breaks gauge invariance. The electron system with such an external field
defines an ensemble that allows for a nonzero superconducting order
parameter, with $\langle\Lambda\rangle$ (and $\langle{\cal G}\rangle$) to 
be determined self-consistently.\cite{Bogoliubov_footnote} 

The Green function ${\tilde G}_{\langle\Lambda\rangle}(x-y)
\equiv{\tilde G}(x-y)$ will play a central role 
in our description of the superconducting phase transition, so we explicitly
give it here. It is convenient to express the matrix elements of
${\tilde G}$ in terms of a normal-state Green function $g$.
In frequency-wavenumber space, with a four-vector 
$k = (\omega_n,{\bf k})$, we define 
\bml
\label{eqs:2.18}
\be
g_{\sigma}^{-1}(k) = i\omega_{n} - \xi_{{\bf k},\sigma}
                       -\Sigma_{\sigma}(k)\quad,  
\label{eq:2.18a}
\ee
with
\be
\xi_{{\bf k},\sigma} = \frac{{\bf k}^2}{2m_{\rm e}} - \mu 
   - \sigma\delta\quad,  
\label{eq:2.18b}
\ee
\eml
where $\sigma = (\uparrow,\downarrow) \equiv (+,-)$ is the spin projection
index, and $\delta = \Gamma_{\rm t}m$ reflects the Stoner splitting
of the Fermi surface in the ferromagnetic phase. We also define a function
\be
D(k) = g_{\uparrow}^{-1}(k)\,g_{\uparrow}^{-1}(-k) 
       + \Delta(k)\,\Delta^{+}(k)\quad.
\label{eq:2.19}
\ee
Finally we note the symmetry relations,
\bml
\label{eqs:2.20}
\bea
\Delta(k) &=& -\Delta(-k)\quad,  
\label{eq:2.20a}\\
\Delta^{+}(k) &=& -\Delta(-k)^{*}\quad,
\label{eq:2.20b}
\eea
\eml%
with the asterisk denoting a complex conjugate. The matrix inversion problem
for ${\tilde G}$ decouples into one $2\times 2$ problem and two $1\times 1$
problems, and we obtain
\bml
\label{eqs:2.21}
\bea
{\tilde G}_{11}(k) &=& g_{\uparrow}^{-1}(-k)/D(k)\quad,  
\label{eq:2.21a}\\
{\tilde G}_{13}(k) &=& \Delta(k)/D(k)\quad,  
\label{eq:2.21b}\\
{\tilde G}_{22}(k) &=& g_{\downarrow}(k)\quad,  
\label{eq:2.21c}
\eea
\eml%
Symmetry gives
\bml
\label{eqs:2.22}
\bea
{\tilde G}_{33}(k) &=& -G_{11}(-k)\quad,
\label{eq:2.22a}\\
{\tilde G}_{44}(k) &=& -{\tilde G}_{22}(-k)\quad,  
\label{eq:2.22b}\\
{\tilde G}_{31}(k) &=& -{\tilde G}_{13}(k)^{*}\quad.  
\label{eq:2.22c}
\eea
\eml%
All other Green's functions are identically equal to zero.

Finally, the inverse magnetic susceptibility 
$\chi^{-1}_{\langle\Lambda\rangle}\equiv\chi^{-1}$
is needed. Physically, $\chi$ represents the normalized magnetic 
susceptibility of a `reference ensemble' of electrons that are subject 
to an external magnetic field equal to $\delta$, and whose self energy 
operator is given by $\Sigma_{\sigma}$. General symmetry arguments show 
that $\chi^{-1}$ has the form
\bml
\label{eqs:2.23}
\bea
\chi^{-1}_{ij}(k) &=& \delta_{ij}\left[\delta_{i3}\chi^{-1}_{\rm L}(k)
                    + (1-\delta_{i3})\chi^{-1}_{\rm T,+}(k) \right]
\nonumber\\
&& \hskip 20pt   +(\delta_{12} - \delta_{21})\,\chi^{-1}_{\rm T,-}(k)\quad,
\label{eq:2.23a}
\eea
with the subscripts $L$ and $T$ denoting longitudinal and transverse 
components, respectively, and the transverse inverse susceptibility has both
diagonal and off-diagonal pieces. Notice that the latter are nonzero
only in a phase with a nonvanishing magnetization (or in the presence of
an external magnetic field). For later reference we give the explicit
expressions that result from our Gaussian approximation for the magnetic
fluctuations. With Eqs. (\ref{eqs:2.15}), (\ref{eqs:2.13}), and
(\ref{eq:2.12b}) we obtain
\bea
\chi^{-1}_{\rm L}(k) &=& 1 + \Gamma_{\rm t}
                         \int_q \left[{\tilde G}_{11}(q){\tilde G}_{11}(q-k)
                         \right.
\nonumber\\
&& \left. + {\tilde G}_{22}(q){\tilde G}_{22}(q-k)
 - {\tilde G}_{31}(q){\tilde G}_{13}(q-k) \right]\quad,
\nonumber\\
\label{eq:2.23b}\\
\chi^{-1}_{\rm T,+}(k) &=& 1 + \Gamma_{\rm t}\int_q {\tilde G}_{11}(q)\left[
                          {\tilde G}_{22}(q+k) + {\tilde G}_{22}(q-k)\right]
                                   \ ,
\nonumber\\
\label{eq:2.23c}\\
\chi^{-1}_{\rm T,-}(k) &=& i\Gamma_{\rm t}\int_q {\tilde G}_{11}(q)\left[
                          {\tilde G}_{22}(q+k) - {\tilde G}_{22}(q-k)\right]
                                   \quad.
\nonumber\\
\label{eq:2.23d}
\eea
\eml%
where $\int_{q} = T\sum_{n}\int d{\bf q}/(2\pi)^{d}$. More explicit
expressions for these quantities will be discussed in 
Sec.\ \ref{subsec:III.D} below.

\section{Saddle-Point Theory for Superconductivity due to Magnetic Fluctuations}
\label{sec:III}

\subsection{Saddle-point solution}
\label{subsec:III.A}

The next step is to solve the field theory defined by the 
Eqs.\ (\ref{eqs:2.17}) in a saddle-point approximation. Notice that there
is only one fluctuating field left, namely ${\cal G}$. For fixed 
$\langle\Lambda\rangle$ we therefore have only one saddle-point
condition,
\be
\delta A/\delta{\cal G}\vert_{\rm sp} = 0\quad.
\label{eq:3.1}
\ee
and we will be looking for a translationally invariant saddle-point solution,
${\cal G}_{\rm sp}(x,y) = {\cal G}_{\rm sp}(x-y) \approx \langle{\cal G}\rangle
(x-y)$ (cf., however, the remarks in Sec.\ \ref{sec:V} below). 
Equation\ (\ref{eq:3.1}) yields an expression for the normal self energies,
Eq.\ (\ref{eq:2.13e}),
\bml
\label{eqs:3.2}
\bea
\Sigma_{\uparrow}(x-y) &=& - \langle\delta\Lambda_{11}\rangle_{\Lambda}(x-y)
\nonumber\\
&& -\Gamma_{\rm s}\delta(x-y)\sum_{\sigma}G_{\sigma}(x-y) \quad,
\label{eq:3.2a}\\
\Sigma_{\downarrow}(x-y) &=& - \langle\delta\Lambda_{22}\rangle_{\Lambda}(x-y)
\nonumber\\
&& -\Gamma_{\rm s}\delta(x-y)\sum_{\sigma}G_{\sigma}(x-y) \quad,
\label{eq:3.2b}
\eea
with $G_{\sigma}$ the normal Green functions, Eq.\ (\ref{eq:2.13d}). For
the anomalous self energies, Eq.\ (\ref{eq:2.13c}), on obtains 
\bea
\Delta^+(x-y) = -\langle\delta\Lambda_{13}\rangle_{\Lambda}(x-y) \quad,
\label{eq:3.2c}\\
\Delta(x-y) = -\langle\delta\Lambda_{31}\rangle_{\Lambda}(x-y) \quad,
\label{eq:3.2d}
\eea
\eml%
Here we have defined a `$\Lambda$-ensemble' that is characterized by a
statistical weight given by the integrand in the last term in
Eq.\ (\ref{eq:2.17b}), with ${\cal G}$ replaced by 
${\cal G}_{\rm sp} \approx \langle{\cal G}\rangle$,
\bml
\label{eqs:3.3}
\be
\langle \ldots \rangle_{\Lambda} = \int D[{\delta\Lambda}]\ \ldots\ 
   e^{B[\delta\Lambda]} /\int D[{\delta\Lambda}]\ e^{B[\delta\Lambda]}\quad,
\label{eq:3.3a}
\ee
where
\bea
B[\delta\Lambda] &=& \Tr({\cal G}_{\rm sp}\delta\Lambda)
   + \frac{1}{2}\,\Tr\ln\left[{\tilde G}^{-1}_{\langle\Lambda\rangle
                                                         + \delta\Lambda}
        /{\tilde G}^{-1}_{\langle\Lambda\rangle}\right]
\nonumber\\
&& \hskip 30pt
 -\frac{1}{2}\,
    \Tr\ln\left[\chi^{-1}_{\langle\Lambda\rangle + \delta\Lambda}
        /\chi^{-1}_{\langle\Lambda\rangle}\right]\quad.
\label{eq:3.3b}
\eea
\eml%

In order to make the Eqs.\ (\ref{eqs:3.2}) more explicit, we now calculate
$\langle\delta\Lambda\rangle_{\Lambda}$ in Gaussian approximation. Expanding
the `action' $B[\delta\Lambda]$ to second order in $\delta\Lambda$, and
performing the Gaussian integral, we obtain
\be
\langle\delta\Lambda\rangle_{\Lambda} = 2{\tilde G}^{-1}\left(
   {\cal G}_{\rm sp}^{\rm T} - \frac{1}{2}\,{\tilde G}\right){\tilde G}^{-1}
     \quad.
\label{eq:3.4}
\ee
Here we have neglected the contribution of the last term in $B[\delta\Lambda]$,
since it represents corrections to mean-field superconductivity due to
magnetic fluctuations. In agreement with our general philosophy of keeping
only leading fluctuation effects, we drop these terms. Inserted in 
Eqs. (\ref{eqs:3.2}), Eq.\ (\ref{eq:3.4}) provides a relation between the
self energies and the saddle-point Green function. The free energy in
saddle-point approximation, $F_{\rm sp} = -T\ln Z_{\rm sp}$ now reads
\bea
F_{\rm sp} &=& \frac{1}{2}V\Gamma_{\rm t}m^2 
    - 2T\Gamma_{\rm s}\int dx\left(\sum_{\sigma}
  G_{\sigma}(x,x)\right)^2
\nonumber\\ 
&&    - T\Tr\left(\langle\Lambda\rangle\,{\cal G}_{\rm sp}
   \right)
- \frac{T}{2}\,\Tr\ln{\tilde G}^{-1}
   + \frac{T}{2}\,\Tr\ln\chi^{-1} 
\nonumber\\
&&   - T\ln\int D[\delta\Lambda]\ e^{B[\delta\Lambda]}\quad,
\label{eq:3.5}
\eea
and it depends on only one independent parameter, e.g., $\langle\Lambda\rangle$.

\subsection{Minimization of the free energy, and strong-coupling equations}
\label{subsec:III.B}

As the last step in our formal development we now require that the physical
value of $\langle\Lambda\rangle$ minimizes the saddle-point free energy.
The condition
\be
\delta F_{\rm sp}/\delta\langle\Lambda\rangle = 0\quad,
\label{eq:3.6}
\ee
leads to 
\bml
\label{eqs:3.7}
\be
\langle\delta\Lambda\rangle_{\Lambda} = {\tilde G}^{-1}\left(\chi\left
   \vert\frac{\delta}{\delta\langle\Lambda\rangle}\right\vert\chi^{-1}
     \right) {\tilde G}^{-1}\quad.
\label{eq:3.7a}
\ee
Here we have used Eq.\ (\ref{eq:3.4}), and a symbolic scalar product notation
\bea
\left(\chi\left\vert\frac{\delta}{\delta\langle\Lambda\rangle}\right\vert
                 \chi^{-1}\right)
   &=& \int dx\,dy\sum_{ij}\chi_{ij}(x-y)\,\frac{\delta}
                                              {\delta\langle\Lambda\rangle}
\nonumber\\
&&\hskip 50pt \times\chi^{-1}_{ji}(y-x)\quad.
\label{eq:3.7b}
\eea
\eml%
Notice that $\langle\delta\Lambda\rangle_{\Lambda}$, ${\tilde G}^{-1}$,
and $\langle\Lambda\rangle$ are still $4\times 4$ matrices in spinor space
(as well as functions of space-time), while $\chi$ is a scalar in spinor space.
Again in agreement with our general
philosophy, we have neglected the contribution of the last term in
Eq.\ (\ref{eq:3.5}), which would lead to superconducting order-parameter 
fluctuation corrections to mean-field superconductivity. Equations
(\ref{eqs:3.2}) and Eq.\ (\ref{eq:3.7a}) now form a closed set of equations
for the various self energies. Due to their matrix nature, they are
complicated, and for our purposes we restrict ourselves to the linearized
gap equations. That is, we evaluate Eqs.\ (\ref{eq:3.7a}) and (\ref{eqs:3.2})
only to linear order in the gap parameter $\Delta$. This procedure is
sufficient for deriving the superconducting phase boundary in both the
paramagnetic and the ferromagnetic phases. We find
\bml
\label{eqs:3.8}
\bea
\Delta(k) &=& 2\Gamma_{\rm t}\int_q \chi_{\rm L}(k-q)\,\Delta(q)\,
               \vert g_{\uparrow}(q)\vert^2\quad,
\nonumber\\
\label{eq:3.8a}\\
\Sigma_{\uparrow}(k) &=& -\Gamma_{\rm s}n + 2\Gamma_{\rm t}\int_q \left[
   \chi_{\rm L}(k-q)\,g_{\uparrow}(q) \right.
\nonumber\\
&&\hskip 65pt \left. + 2\chi_{\rm T,+}(k-q)\,g_{\downarrow}(q)\right]\quad,
\label{eq:3.8b}\\
\Sigma_{\downarrow}(k) &=& -\Gamma_{\rm s}n + 2\Gamma_{\rm t}\int_q \left[
   \chi_{\rm L}(k-q)\,g_{\downarrow}(q) \right.
\nonumber\\
&&\hskip 65pt \left. + 2\chi_{\rm T,+}(k-q)\,g_{\uparrow}(q)\right]\quad,
\label{eq:3.8c}
\eea
\eml%
with $g_{\uparrow}$ and $g_{\downarrow}$ given by Eqs.\ (\ref{eqs:2.18}).
A comparison with Eqs.\ (\ref{eqs:A.10}) shows that Eqs.\ (\ref{eqs:3.8})
are an obvious generalization of the conventional Eliashberg equations to
the case of superconductivity mediated by spin fluctuations, with the
magnetic susceptibility essentially replacing the phonon propagator, 
and the inequivalency of the two spin projections taken into account.

The susceptibilities that appear in the Eqs.\ (\ref{eqs:3.8}) are given
explicitly by Eqs.\ (\ref{eqs:2.23}). It is obvious from our derivation,
however, that these explicit expressions are just a particular approximation
for the exact susceptibilities, and that the latter will appear in the theory
if one takes loop corrections into account. We will therefore consider the 
$\chi$ as physical susceptibilities that we model in the next subsections, 
using general physical arguments. We also note that
in the paramagnetic phase, $\chi_{\rm L}=\chi_{\rm T}$. However,
in the ferromagnetic phase the two susceptibilities are fundamentally 
different, as we will discuss below.

Finally, we have yet to discuss the magnetization as a function of the 
control parameters that drive the magnetic transition. This is obtained
by minimizing the free energy with respect to $m$, in order to find
the physical values of the magnetization,
\be
\delta F_{\rm sp}/\delta m = 0\quad.
\label{eq:3.9}
\ee
A detailed theory of
the itinerant ferromagnetic phase transition at low temperatures has been
given elsewhere.\cite{us_1st_order} 
In this paper, where we are interested in the 
superconducting transition, we restrict ourselves to treating the magnetization
in mean-field theory. That is, we neglect the fluctuation effects 
that are represented by the last two terms on the right-hand side of
Eq.\ (\ref{eq:3.5}). We then obtain the magnetic mean-field
equation of state in the form
\be
m = \frac{T}{2V}\,\Tr (\gamma_3{\tilde G})\quad.
\label{eq:3.10}
\ee
We will discuss this expression explicitly in Sec.\ \ref{subsubsec:III.D.2}
below.

\subsection{Gap equation for p-wave superconductivity}
\label{subsec:III.C}

So far the symmetry of the superconducting order parameter has not been
specified. To do so, we expand $\Delta (k)$ in Legendre polynomials,
\be
\Delta (k) = \Delta (i\omega_n,\vert{\bf k}\vert,\cos\theta) 
           = \sum_{l=0}^{\infty}\Delta^{(l)}(i\omega_n,\vert{\bf k}\vert)\,
             P_l(\cos\theta)\quad,
\label{eq:3.11}
\ee
where $\theta$ is the azimuthal angle on the spherical Fermi surface.
Since $\Delta(k) = -\Delta(-k)$, the $\Delta^{(l)}$ are odd (even) functions
of the frequency for even (odd) values of $l$.
If we insert this expansion in Eq.\ (\ref{eq:3.8a}), we see that in
the static limit only odd values of $l$ contribute. Of the contributions 
with odd angular momenta, $l=1$ will be the strongest for phase space 
reasons.\cite{p_wave_footnote} We thus specialize to the
p-wave case, and use the notation $\Delta^{(1)} \equiv \Delta$. We further
split off the frequency dependence of the normal self energy and write, as
in Eliashberg theory,\cite{AGD}
\bml
\label{eqs:3.12}
\be
G_{\rm n,\uparrow}^{-1}(q) = i\omega_n Z_{\uparrow}(q) 
   - {\tilde \xi}_{{\bf q},\uparrow}\quad,
\label{eq:3.12a}
\ee
with
\be
Z_{\uparrow}(q) = 1 - \left[\Sigma_{\uparrow}(i\omega_n,{\bf q}) 
                            - \Sigma_{\uparrow}(0,{\bf q})\right]/i\omega_n
                                    \quad,
\label{eq:3.12b}
\ee
and
\be
{\tilde \xi}_{{\bf q},\uparrow} = \xi_{{\bf q},\uparrow} 
     + \Sigma_{\uparrow}(0,{\bf q})\quad.
\label{eq:3.12c}
\ee
\eml%
The normal state Green functions in Eq.\ (\ref{eq:3.8a}) then pin the
momentum to the up-spin Fermi surface, and we obtain for
$\Delta^{(1)}(i\omega_n,k_{\rm F}^{\uparrow}) \equiv \Delta(i\omega_n)$
the equation
\bml
\label{eqs:3.13}
\be
\Delta(i\omega_n)=\pi T\sum_{i\omega_m} d^{\,1}_{\rm L}(i\omega_n-i\omega_m)\,
                   \Delta(i\omega_m)/\vert{\tilde\omega}_m\vert\quad,
\label{eq:3.13a}
\ee
where $i{\tilde\omega}_m = i\omega_m Z_{\uparrow}(i\omega_m)$.
In the same way we obtain an equation for $Z_{\uparrow}$,
\bea
\omega_n \left[1 - Z_{\uparrow}(i\omega_n)\right] &=& -\pi T\sum_{i\omega_m}
   \sgn(\omega_m) \left[d^{\,0}_{\rm L}(i\omega_n - i\omega_m)\right.
\nonumber\\
&&\hskip 20pt \left.   + 2d^{\,0}_{\rm T}(i\omega_n - i\omega_m)\right]\quad.
\label{eq:3.13b}
\eea
\eml%
The kernels in these integral equations are given by
\bml
\label{eqs:3.14}
\bea
d_{\rm L}^{\,1}(i\Omega_n)&=&\Gamma_{\rm t}N_{\rm F}^{\uparrow}
  \int_0^2 dx\,x (1-x^2/2)\,\chi_{\rm L}(xk_{\rm F}^{\uparrow},i\Omega_n)\quad,
\nonumber\\
\label{eq:3.14a}\\
d_{\rm L}^{\,0}(i\Omega_n)&=&\Gamma_{\rm t}N_{\rm F}^{\uparrow}
   \int_0^2 dx\,x \,\chi_{\rm L}(xk_{\rm F}^{\uparrow},i\Omega_n)\quad,
\label{eq:3.14b}\\
d_{\rm T}^{\,0}(i\Omega_n)&=&\Gamma_{\rm t}N_{\rm F}^{\uparrow}
   \int_{x_-}^{x_+} dx\,x\,\chi_{\rm T}(xk_{\rm F}^{\uparrow},i\Omega_n)\quad.
\label{eq:3.14c}
\eea
\eml%
Here $N_{\rm F}$ and $N_{\rm F}^{\uparrow}$ are the density of states per
spin at the Fermi level in the paramagnetic phase, and the density of states
at the Fermi level for up-spin electrons in the ferromagnetic phase,
respectively, and
$x_{\pm} = 1 \pm k_{\rm F}^{\downarrow}/k_{\rm F}^{\uparrow}$.

\subsection{Explicit model for the longitudinal spin susceptibility}
\label{subsec:III.D}

For an explicit evaluation of our strong-coupling theory we need the
longitudinal and transverse magnetic susceptibilities as input. For
the determination of the superconducting $T_{\rm c}$, or the phase
diagram, it suffices to know these susceptibilities in the normal-metal 
state. This is equivalent to the need of the phonon spectrum
as input in ordinary Eliashberg theory. As in the latter case, one
would ideally want to use experimental data as input, but this
information is not available for the materials of interest. We
therefore resort to theoretical modeling, with the aim of developing
theoretical expressions that correctly reflect the relevant qualitative 
physical effects. Since the latter are different in the paramagnetic
and ferromagnetic phases, respectively, we will treat these two cases
separately.

\subsubsection{Paramagnetic phase}
\label{subsubsec:III.D.1}

In the paramagnetic phase, spin rotational invariance implies 
$\chi_{\rm L} = \chi_{\rm T} \equiv \chi$. Equation (\ref{eq:2.23b}) gives an
explicit expression which, after substituting the Green functions
${\tilde G}$ and using Eqs.\ (\ref{eqs:3.8}), results in an integral
equation for $\chi_{\rm L}$. This is the result of a particular
approximation we made in Sec.\ \ref{sec:II}, that integrated out
the magnetic fluctuations in a Gaussian approximation. This
particular procedure results in a fairly elaborate treatment of
the denominator of $\chi$, which takes into account the self energies
of the Green functions, while the numerator is approximated by a
constant, neglecting its wavenumber and frequency dependence. A slightly
different Gaussian approximation was derived in Ref.\ \onlinecite{us_fm_mit}
by expanding about a Stoner saddle point within a theory focussing purely on 
magnetism. This approximation neglects the self energy $\Sigma_{\sigma}$ in
the Green function $g_{\sigma}$, Eq.\ (\ref{eq:2.18a}), but keeps the
wavenumber and frequency dependence of the numerator of $\chi$ at the
same level as that of the denominator. 
In the static, long-wavelength limit, and not too far from the magnetic 
transition, both of these zero-loop approximations reduce to the Landau 
expression\cite{normalization_footnote}
\be
\chi_{\rm L}^{(0)}({\bf k},i0) = \chi_{\rm T}^{(0)}({\bf k},i0)
                        = \frac{1-t}{t + b({\bf k}/2k_{\rm F})^2}\quad.
\label{eq:3.15}
\ee
Here $t = 1-2N_{\rm F}\Gamma_{\rm t}$ is the dimensionless distance from 
the magnetic critical point in mean-field approximation. The value of the
coefficient $b$ of course depends on the model. If one neglects the self
energy in the Green function in Eq.\ (\ref{eq:2.23b}), or in the 
approximation of
Ref.\ \onlinecite{us_fm_mit}, $b=1/3$, but in general there is no reason 
to prefer this value of $b$ over any other value of order unity.

We have used both the model from Ref.\ \onlinecite{us_fm_mit}, and the
Landau model, Eq.\ (\ref{eq:3.15}), to calculate the coupling constants
$d^{\,0,1}_{\rm L,T}$, Eqs.\ (\ref{eqs:3.14}), and the resulting 
superconducting $T_{\rm c}$ in a McMillan approximation, see Sec.\ \ref{sec:IV}
below. We have found that for reasonable values of $b$ the results are 
qualitatively the same. In what follows, we will use the Landau approximation,
Eq.\ (\ref{eq:3.15}), with $b$ as a free parameter. We emphasize that this
approximation correctly reflects the long-wavelength behavior of the magnetic 
susceptibility in the paramagnetic phase.
The free parameter $b$ just reflects deviations from the
free electron model that one would expect in any real material.

\subsubsection{Ferromagnetic phase, zero-loop order}
\label{subsubsec:III.D.2}

In the ferromagnetic phase, any reasonable zero-loop approximation for
the susceptibilities yields again a Landau expression for the longitudinal
static susceptibility in the long-wavelength limit. 
We evaluate Eq.\ (\ref{eq:2.23b}), again neglecting the self energy, and 
find\cite{normalization_footnote}
\be
\chi_{\rm L}^{(0)}({\bf k},i0) = \frac{1-t}{a_{\rm L}\vert t\vert
                         + b_{\rm L}({\bf k}/2k_{\rm F})^2}\quad,
\label{eq:3.16}
\ee
where we have introduced two free parameters, $a_{\rm L}$ and $b_{\rm L}$. 
In ordinary Landau theory, $a_{\rm L} = 2$. However, if
one keeps the electron number density fixed, as is the case experimentally,
rather than the chemical potential, then in general $a_{\rm L}\neq 2$.
From the point of view
of Landau theory, this can be understood as follows. Since
$t = 1 - 2N(\mu)\Gamma_{\rm t}$, and the chemical potential $\mu$
depends on the magnetization, the coefficient of the magnetization squared
in the Landau free energy depends itself on the magnetization, which
causes the deviation of the prefactor of $\vert t\vert$ from the usual
value of $2$. 

The actual value of $a_{\rm L}$ is model dependent, and depends in
particular on the equation of state, since the latter determines the
relation between $\vert t\vert$ and the magnetization. If we again
neglect the self energy in the Green function on the right-hand side
of Eq.\ (\ref{eq:3.10}), the latter yields
\bml
\label{eqs:3.17}
\be
6\pi^2\delta = \Gamma_{\rm t}(2m_{\rm e})^{3/2}\left[(\mu + \delta)^{3/2}
               - (\mu - \delta)^{3/2}\right]\quad.
\label{eq:3.17a}
\ee
This needs to be supplemented by the condition that the electron 
number density $n_{\rm e}$ is fixed, which is
\be
6\pi^2 n_{\rm e} = (2m_{\rm e})^{3/2}\left[(\mu + \delta)^{3/2}
           + (\mu - \delta)^{3/2}\right]\quad.
\label{eq:3.17b}
\ee
\eml%
This determines the chemical potential $\mu$ as a function of the
magnetization. Equations (\ref{eqs:3.17}) together determine the
relation between $\delta$ and $\vert t\vert$.

From Equations (\ref{eqs:3.17}) and Eq.\ (\ref{eq:2.23b}) one obtains
$a_{\rm L} = 1/2$. Different Gaussian approximations for $\chi$ yield
different values. For instance, Ref.\ \onlinecite{us_fm_mit} in
conjuction with Eqs.\ (\ref{eqs:3.17}), and putting 
$\Gamma_{\rm s} = \Gamma_{\rm t}$, yields $a_{\rm L} = 5/4$, and
so does the RPA expression used by Fay and Appel.\cite{FayAppel}
In order to be able to compare with the latter's calculation of $T_{\rm c}$,
we adopt $a_{\rm L} = 5/4$ in what follows. Other values of $O(1)$
do not lead to qualitatively different results.

In the transverse channel, a Ward identity guarantees that $\chi$ diverges
in the limit of small wavenumbers and frequencies. This reflects the
existence of magnetic Goldstone modes or magnons. Our mean-field equation
of state, Eq.\ (\ref{eq:3.10}), or Eq.\ (\ref{eq:3.17a}), is consistent
with our Gaussian treatment of the magnetic fluctuations,  
Eqs.\ (\ref{eqs:2.23}), in the sense that they form a conserving approximation
that correctly reflects the Goldstone modes. Inverting the susceptibility
tensor given by Eqs.\ (\ref{eqs:2.23}), dropping the self energy in the
Green functions, and using Eq.\ (\ref{eq:3.17a}), we find for
$\chi_{11}(k) \equiv \chi_{\rm T,+}(k)$ in Gaussian 
approximation,\cite{normalization_footnote} 
\bml
\label{eqs:3.18}
\bea
\chi_{\rm T,+}^{(0)}({\bf k},i\Omega_n) &=& \frac{\delta/4\epsilon_{\rm F}}
                                             {1 - t}\,
   \left(\frac{1}{i\Omega_n/4\epsilon_{\rm F} + (\delta/2\epsilon_{\rm F})
                  b_{\rm T}({\bf k}/2k_{\rm F})^2}   \right.
\nonumber\\
&& \left. - \frac{1}{i\Omega_n/4\epsilon_{\rm F} - (\delta/2\epsilon_{\rm F})
                  b_{\rm T}({\bf k}/2k_{\rm F})^2}   \right)\quad,
\nonumber\\
\label{eq:3.18a}
\eea
and for $\chi_{12}(k) \equiv \chi_{\rm T,-}(k)$,
\bea
\chi_{\rm T,-}^{(0)}({\bf k},i\Omega_n)\hskip -1pt &=& \hskip -1pt 
              \frac{-i\delta/4\epsilon_{\rm F}}{1 - t}\hskip -2pt
   \left(\hskip -1pt\frac{1}{i\Omega_n/4\epsilon_{\rm F} 
        + (\delta/2\epsilon_{\rm F}) b_{\rm T}({\bf k}/2k_{\rm F})^2}   \right.
\nonumber\\
&& \left. + \frac{1}{i\Omega_n/4\epsilon_{\rm F} - (\delta/2\epsilon_{\rm F})
                  b_{\rm T}({\bf k}/2k_{\rm F})^2}   \right)\quad.
\nonumber\\
\label{eq:3.18b}
\eea
\eml%
Here we give the dynamical susceptibility, since we will need it later.
In our approximation, $b_{\rm T} = 1/3$, but in general it is some parameter
of order unity, and in general it will be different from $b_{\rm L}$ in the 
ferromagnetic phase.

\subsubsection{Ferromagnetic phase, one-loop order}
\label{subsubsec:III.D.3}

\begin{figure}[ht]
\epsfxsize=65mm
\centerline{\epsffile{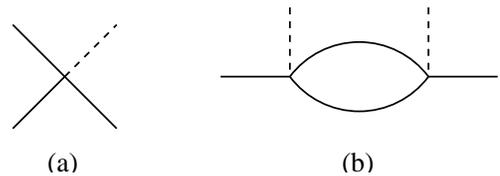}}
\vskip 5mm
\caption{(a) A contribution to the quartic magnetization vertex, with the
 dashed line representing an average magnetization, and the solid line a
 magnetization fluctuation. (b) A resulting contribution to the magnetic
 susceptibility.}
\vskip 0mm
\label{fig:2}
\end{figure}

In contrast to the situation in the paramagnetic phase, in the ferromagnetic
phase the zero-loop approximation for the longitudinal susceptibility, 
Eq.\ (\ref{eq:3.16}) is {\em not} qualitatively correct. The reason for
this is as follows. In a Heisenberg ferromagnet, the transverse spin waves
or magnons couple to the longitudinal spin fluctuations, and hence contribute
to the longitudinal susceptibility $\chi_{\rm L}$.\cite{Ma} At nonzero
temperature, and for dimensions $d<4$, this coupling actually leads to a
magnetic susceptibility that diverges everywhere on the coexistence 
curve.\cite{BrezinWallace} More generally, and more importantly for our
purposes, it causes the zero-loop or Landau expression for the longitudinal 
susceptibility, Eq.\ (\ref{eq:3.16}), to be qualitatively incorrect in the 
ferromagnetic phase. Going to one-loop order, on the other hand, 
produces a functional
form of $\chi_{\rm L}$ that is exact at small wavenumbers. This one-loop
contribution can be calculated in a variety of ways. Suppose we expand the 
action, Eq.\ (\ref{eq:2.12a}), in powers of ${\bf M}$, to quartic order.
Then each of the four ${\bf M}$ fields in the quartic vertex can represent
either an average magnetization, or a magnetization fluctuation, see
Eq.\ (\ref{eq:2.13a}). This leads to a one-loop contribution to the
two-point ${\bf M}$-vertex, or the inverse magnetic susceptibility, that
has the structure shown in Fig.\ \ref{fig:2}.
Equivalently, an expansion in powers of $\delta{\bf M}$ leads to a cubic
vertex, and to a diagram with the same structure as in Fig.\ \ref{fig:2}(b).
Alternatively, one can use a nonlinear sigma-model to derive the same
contribution, as was done in Ref.\ \onlinecite{us_letter}. Either way
one obtains for the one-loop contribution to the longitudinal
susceptibility\cite{normalization_footnote}
\bea
\chi_{\rm L}^{(1)}(k) &=& \frac{2\Gamma_{\rm t}}{\delta^2}
   \,\chi_{\rm L}^{(0)}(k)\int_q \left[\chi^{(0)}_{\rm T,+}(k-q)\,
                                        \chi^{(0)}_{\rm T,+}(q)\right.
\nonumber\\
&&+ \left. \chi^{(0)}_{\rm T,-}(k-q)\,\chi^{(0)}_{\rm T,-}(q)\right]\,
                      \chi_{\rm L}^{(0)}(k)\quad.
\label{eq:3.19}
\eea
Here $\chi^{(0)}_{\rm T,\pm}$ are the zero-loop transverse susceptibility
tensor elements, Eqs.\ (\ref{eqs:3.18}), and $\chi_{\rm L}^{(0)}$ is the 
zero-loop longitudinal susceptibility, Eq.\ (\ref{eq:3.16}). The integral in
Eq.\ (\ref{eq:3.19}) leads to $\chi_{\rm L}({\bf k},i0) \propto
\vert{\bf k}\vert^{d-4}$ for the leading wavenumber dependence of
$\chi_{\rm L}$ in $d<4$, which is the exact leading behavior for
small wavenumbers.\cite{k^{d-4}_footnote} We note that there is no analog 
of the diagram shown in Fig.\ \ref{fig:2}(b) in the paramagnetic phase, as
there is no cubic magnetization vertex. It is therefore reasonable to
use the zero-loop expression for the longitudinal susceptibility in
the paramagnetic phase, and the one-loop expression given by 
Eq.\ (\ref{eq:3.19}) in the ferromagnetic one.

We now have explicit model expressions, in terms of integrals, for the
magnetic susceptibilities that we need as input for our theory of
superconductivity. In the paramagnetic phase, the simple expression
given in Eq.\ (\ref{eq:3.15}) suffices. In the ferromagnetic phase,
we will use Eq.\ (\ref{eq:3.18a}) for the transverse susceptibility,
and $\chi_{\rm L}^{(0)}({\bf k},i0) + \chi_{\rm L}^{(1)}({\bf k},i0)$,
Eqs.\ (\ref{eq:3.16}) and (\ref{eq:3.19}), with $a_{\rm L} = 5/4$, 
for the longitudinal one. This model contains two parameters of $O(1)$, 
viz. $b_{\rm L}$ and $b_{\rm T}$, and it reflects the exact functional 
form of the susceptibilities in the long-wavelength limit.

\section{McMillan-Type Solution of the Gap Equation}
\label{sec:IV}

The strong coupling equations, Eqs.\ (\ref{eqs:3.13}), have the same
structure as Eliashberg theory for phonon-induced superconductivity.
There are well-developed algorithms for solving these equations. It
is well known that any attempt to quantitatively calculate the critical
temperature requires a numerical solution of the equations, using
detailed information about the kernels as input. In the case of
superconductivity induced by magnetic fluctuations, the latter information
is not available, and we will have to use the schematic model for the
susceptibility that was developed in the last section. We will therefore
calculate the superconducting $T_{\rm c}$ by means of a simple expression
of McMillan or Allen-Dynes type.\cite{AllenDynes} 
Since we are interested in the structure
of the phase diagram, and in particular in the {\em relative} values of
the superconducting $T_{\rm c}$ in the paramagnetic and ferromagnetic
phases, respectively, these approximations will give the desired information
provided we treat the two phases on equal footing. As we have explained in
the last section, this is the case if we use the susceptibility model
represented by Eqs.\ (\ref{eq:3.15}), (\ref{eq:3.18a}), and (\ref{eq:3.19}),
respectively. For a rough order-of-magnitude estimate of the absolute
value of $T_{\rm c}$, see Sec.\ \ref{subsec:V.A} below.

\subsection{$T_{\rm c}$ formula}
\label{subsec:IV.A}

McMillan's approximate solution of the linearized strong-coupling equations
results in a critical temperature
\be
T_{\rm c} = T_0(t)\,e^{-(1 + d_{\rm L}^0 + 2d_{\rm T}^0)/d_{\rm L}^1}\quad.
\label{eq:4.1}
\ee
Here $d_{\rm L,T}^0$ and $d_{\rm L}^1$ are the zero-frequency values of the
functions defined in Eqs.\ (\ref{eqs:3.14}), and $T_0(t)$ is a temperature
scale that we will define and discuss below. $d_{\rm T}^0$, which involves
an integral over a zero-loop susceptibility, is easily calculated analytically.
For $d_{\rm L}^{0,1}$ we proceed as follows. The momentum integral of the
convolution in Eq.\ (\ref{eq:3.19}) can be done analytically. For the external
legs in that expression, we replace Eq.\ (\ref{eq:3.16}) by a step function
$\Theta(\vert{\bf k}\vert - 2k_{\rm F}x_{\rm c})$,
with $x_{\rm c} = \sqrt{5\vert t\vert/4b_{\rm L}}$. 
The momentum integration in Eq.\ (\ref{eq:3.14b}) can 
then also be performed analytically. This leaves us with the frequency
summation in Eq.\ (\ref{eq:3.19}), which needs to be done numerically.
We obtain
\bml
\label{eqs:4.2}
\bea
d_{\rm L}^0 &=& \frac{8\pi}{b_{\rm T}^2}\,
                \frac{1}{(\delta/\epsilon_{\rm F})^2}\,\frac{1}{\delta_+}\,
                \frac{T_{\rm c}}{\delta}\,\left[\pi y_c
                  + 2\sum_{n=1}^{N} f(\Omega_n,y_c)\right]\,,
\nonumber\\
\label{eq:4.2a}
\eea
Here $\delta_+ = \sqrt{1 + \delta/\epsilon_{\rm F}}$, $y_c = {\rm Min}(1,x_c)$, 
and
\bea
f(\Omega_m,y) &=& a_m\left[\frac{\pi}{2} + \ln 2 - 
                            \frac{\vert a_m-y\vert}{a_m}\,
                \arctan\left(\frac{a_m}{\vert a_m-y\vert}\right)\right.
\nonumber\\
&&\hskip -50pt - \frac{a_m+y}{a_m}\,\arctan\left(\frac{a_m}{\vert a_m+y\vert}
                                                                      \right)
   - \frac{1}{2}\,\ln\left(1 + \frac{(a_m-y)^2}{a_m^2}\right)
\nonumber\\
&&\hskip -26pt \left. - \frac{1}{2}\,\ln\left(1 + \frac{(a_m+y)^2}{a_m^2}\right)
   + \Theta(y - a_m)\pi\,\frac{y - a_m}{a_m}\right]\ ,
\nonumber\\
\label{eq:4.2b}
\eea
with $a_m = \sqrt{\Omega_m/b_{\rm T}\delta}$.
$N$ in Eq.\ (\ref{eq:4.2a}) is given by $N = \Omega_0/2\pi T_{\rm c}$, with
$\Omega_0$ a frequency cutoff that is on the order of $2b_{\rm T}\delta$.
Equation (\ref{eq:3.14a}) has an additional factor $1-x^2/2$ in the integrand,
which we approximately take into account, in the same spirit as our
approximation for the external legs, by writing 
\bea
d_{\rm L}^1 &=& \frac{8\pi}{b_{\rm T}^2}\,
                \frac{1}{(\delta/\epsilon_{\rm F})^2}\,\frac{1}{\delta_+}\,
                \frac{T_{\rm c}}{\delta}\,\left[\pi y_c^1
                  + 2\sum_{n=1}^{N} f(\Omega_n,y_c^1)\right],
\nonumber\\
\label{eq:4.2c}
\eea
\eml%
with $y_c^1 = {\rm Min}(x_c,1/2\sqrt{2})$.

Finally, we need to specify the temperature scale $T_0(t)$. Following
Refs.\ \onlinecite{FayAppel,BrinkmanEngelsberg}, we use the prefactor of
$\vert t\vert$ in Eqs.\ (\ref{eq:3.15}) and (\ref{eq:3.16}) as a rough
measure of the magnetic excitation energy,
\be
T_0(t) = T_0\,\left[\Theta(t) t + \Theta(-t) 5\vert t\vert/4\right],
\label{eq:4.3}
\ee
with $T_0$ a microscopic temperature scale that is related to the Fermi
temperature, for a free electron model, or a band width, for a band electron
model. This choice of $T_0(t)$ qualitatively reflects the suppression of the
superconducting $T_{\rm c}$ near the ferromagnetic transition due to effective
mass 
effects.\cite{LevinValls,FayAppel,Wang_et_al,RoussevMillis,finite_Tc_footnote}

Equations (\ref{eq:4.1}) - (\ref{eq:4.3}) now form a closed set of
transcendental equations for the superconducting $T_{\rm c}$, which we
have solved by numerical iteration. We discuss the results in the
next subsection.

\subsection{Results}
\label{subsec:IV.B}

We have solved the $T_{\rm c}$ equations (\ref{eq:4.1}) - (\ref{eq:4.3})
for various values of the parameters, and with various implementations of
our approximations. Specifically, we have also used RPA expressions for the
magnetic susceptibilities\cite{BrinkmanEngelsberg,us_fm_mit} instead of
the simple Landau expressions given above. While this makes the calculations
substantially more complicated, the results were qualitatively the same.
We therefore only show results obtained by using the Landau model. We have
also modified the upper frequency cutoff, and again have found only
quantitative effects. Since $T_{\rm c}$ depends exponentially on the
coupling constants, the quantitative effect of changing the cutoff is
substantial. However, the relative values
of $T_{\rm c}$ in the paramagnetic and ferromagnetic phases, respectively,
depend only very weakly on the cutoff. We therefore use 
$\Omega_0 = 2b_{\rm T}\delta$ throughout. Finally, in the ferromagnetic
phase it is illustrative to relate the superconducting $T_{\rm c}$ to the
magnetization in addition to $t$, so we need again the magnetic equation of
state. Within our mean-field model for the magnetization, 
Eqs.\ (\ref{eqs:3.17}) it is useful to introduce a parameter 
$\eta$ via\cite{BrinkmanEngelsberg}
\bml
\label{eqs:4.4}
\be
t = 1 - (1 + 3\eta^2)^{1/3}/(1 + \eta^2/3)\quad.
\label{eq:4.4a}
\ee
The magnetization, in units of $\mu_{\rm B}n$ with $\mu_{\rm B}$ the Bohr
magneton and $n$ the electron number density, is related to $\eta$
by\cite{BrinkmanEngelsberg}
\be
m/\mu_{\rm B}n_{\rm e} = 3\eta(1 + \eta^2/3)/(1 + 3\eta^2)\quad.
\label{eq:4.4b}
\ee
\eml%

Finally, we take into account theoretical\cite{us_1st_order} and 
experimental\cite{experimental_1st_order} evidence that the quantum 
ferromagnetic transition in clean itinerant electron systems is 
of first order at least in some systems. 
This is because fluctuation effects first appear
at one-loop order lead in general to a negative term in the Landau free
energy, i.e., one has a fluctuation-induced first order 
transition.\cite{us_1st_order,us_tbp} A very simple way to take this
into account in the present context is to just impose a minimum value
$t_{\rm min}$ on $\vert t\vert$, which corresponds to a minimum value of the
magnetization in the ferromagnetic phase. 

\begin{figure}[ht]
\epsfxsize=105mm
\centerline{\epsffile{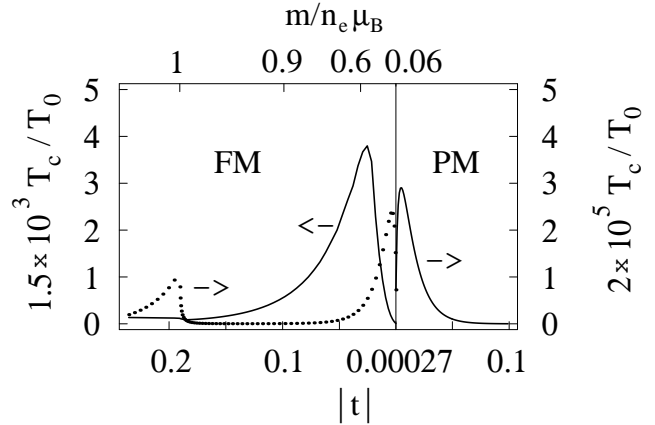}}
\vskip 5mm
\caption{Superconducting $T_{\rm c}$ as a function
 of the distance from the critical point $t$, and the magnetization $m$. The
 solide curve in the PM phase and the dotted curve in the FM phase 
 (right scale) show the zero-loop $T_{\rm c}$ 
 scaled by a factor of 133.3, while the solid curve in the FM phase (left 
 scale) represents the one-loop result. See the text for parameter values
 and further explanation.}
 \vskip 0mm
\label{fig:3}
\end{figure}

\begin{figure}[ht]
\epsfxsize=105mm
\centerline{\epsffile{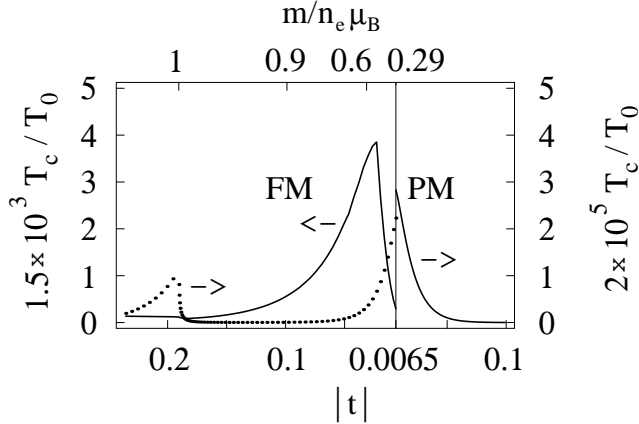}}
\vskip 5mm
\caption{Same as Fig.\ \protect\ref{fig:3}, but with a different value
 of $t_{\rm min}$, see the text.}
\label{fig:4}
\end{figure}

Figure \ref{fig:3} shows the superconducting $T_{\rm c}$ as a function of
$t$ with $t_{\rm min} = 2.66\times 10^{-4}$, which corresponds to a
discontinuity in the magnetization equal to 6\% of the saturation value, 
$b = b_{\rm L} = 0.23$, and $b_{\rm T} = 0.25$. 
Also shown is the magnetization $m$ in the ferromagnetic phase in units of 
the saturation magnetization $\mu_{\rm B}n_{\rm e}$.
$T_{\rm c}$ is measured in units of the characteristic temperature $T_0$
that is given by either the Fermi temperature or the band width, depending
on the model considered. The solid curves show the result in the 
paramagnetic and ferromagnetic phases, respectively, with the former scaled
by a factor of $133.3$ (right-hand scale) compared to the latter (left-hand
scale). The dotted curve in the
ferromagnetic phase (also scaled by a factor of $133.3$, right-hand scale)
represents the result that is obtained upon neglecting the mode-mode
coupling effect. The zero-loop results in the paramagnetic and ferromagnetic
phases are very close to one another, and also very close to the result
of Fay and Appel.\cite{FayAppel} Upon inclusion of the mode-mode coupling
effect at one-loop order, however, the maximum value of $T_{\rm c}$ in the
ferromagnetic phase is enhanced by a factor of more than $100$. As we have
explained above, there is no analogous contribution in the paramagnetic 
phase, so this comparison is physically sensible. We also reiterate that
one should not take the absolute $T_{\rm c}$ values very seriously. However,
the relative comparison we expect to be reliable. We thus find a pronounced
asymmetry between the paramagnetic and ferromagnetic phases. In the case
of UGe$_2$, where the observed masimum $T_{\rm c}$ in the ferromagnetic 
phase is about 500 mK, we thus predict that in the paramagnetic phase no 
superconductivity should be expected above temperatures of at most about 5 mK. 
This is in agreement with the experimentally observed absence of 
superconductivity in the paramagnetic phase. The second maximum in the
Gaussian $T_{\rm c}$ in the FM phase we will discuss below. 
Fig.\ \ref{fig:4} shows the result with the same parameters, except that
$t_{\rm min} = 6.5\times 10^{-3}$, corresponding to a magnetization
discontinuity equal to 29\% of the saturation value. Figures \ref{fig:3}
and \ref{fig:4} correspond to systems with a magnetic phase transition
that is weakly and strongly first order, respectively.

\begin{figure}[ht]
\epsfxsize=105mm
\centerline{\epsffile{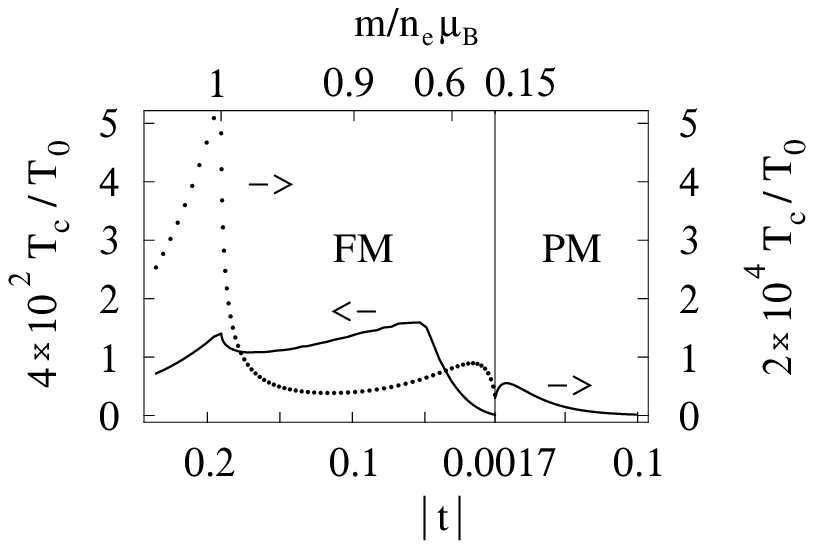}}
\vskip 5mm
\caption{Same as Fig.\ \protect\ref{fig:3}, but for different parameter
 values. See the text for further explanation.}
\vskip 0mm
\label{fig:5}
\end{figure}

\begin{figure}[ht]
\epsfxsize=105mm
\centerline{\epsffile{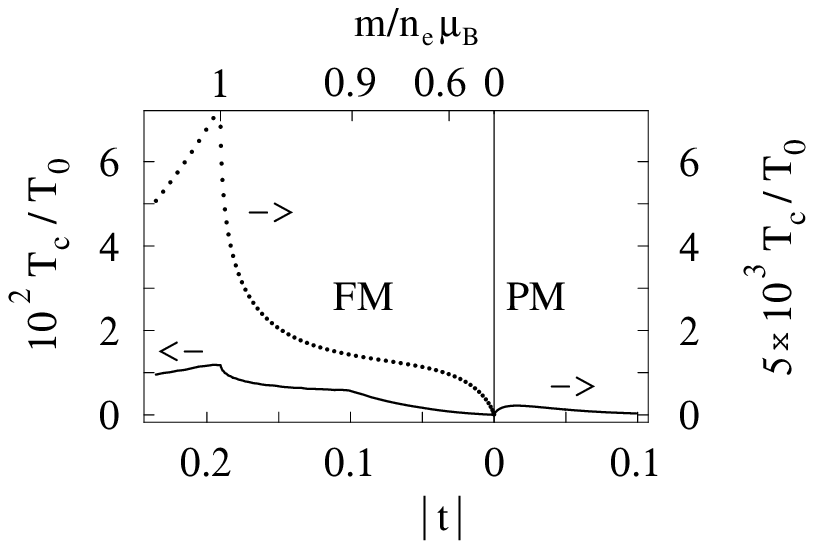}}
\vskip 5mm
\caption{Same as Fig.\ \protect\ref{fig:3}, but for different parameter
 values. See the text for further explanation.}
\vskip 0mm
\label{fig:6}
\end{figure}

We now discuss the effect of varying the parameters $b_{\rm L}$ and $b_{\rm T}$.
With $b = b_{\rm L} = b_{\rm T} = 0.5$ and
$t_{\rm min} = 1.7\times 10^{-3}$ we obtain the result shown in
Fig.\ \ref{fig:5}. $b = b_{\rm L} = b_{\rm T} = 1.0$ and
$t_{\rm min} = 0$ yields Fig.\ \ref{fig:6}. The left and right-hand scales
differ by a factor of $50$ in both figures.
While the (unphysical) zero-loop result in the ferromagnetic
phase is very sensitive to the parameter values, we see that the enhancement
of the (physical) one-loop result over the $T_{\rm c}$ in the paramagnetic
phase is rather robust. However, the position of the maximum of $T_{\rm c}$
changes compared to Fig.\ \ref{fig:3}; in Fig.\ \ref{fig:6} it occurs at the 
point where the magnetization reaches its saturation value, and in 
Fig.\ \ref{fig:5} there are two maxima of about equal hight. The 
reason is as follows. As one
approaches the magnetization saturation point from low magnetization values,
the transverse coupling constant $d_{\rm T}^0$ vanishes, and remains zero in
the saturated region. Effectively, the Heisenberg system turns into the Ising
model discussed in Ref.\ \onlinecite{RoussevMillis}. If the longitudinal
coupling constant $d_{\rm L}^1$ still has a substantial value at that point,
then this leads to an increase in $T_{\rm c}$. This is a very strong effect
in the zero-loop contribution, see Figs.\ \ref{fig:5} and \ref{fig:6}, 
and the effect
qualitatively survives in the one-loop result. If, however, $d_{\rm L}^1$ is
already very small, then $d_{\rm T}^0$ going to zero has only a very small
effect on $T_{\rm c}$, as is the case in Figs.\ \ref{fig:3} and \ref{fig:4},
although the effect is still visible in the zero-loop result. 
Which of these two cases is realized depends on the parameter values. 
This observation provides an explanation for the difference in the observed
phase diagrams in UGe$_2$ and ZrZn$_2$, respectively. UGe$_2$ reaches
its saturation magnetization relatively close to the magnetic phase
transition, at a pressure well above ambient pressure. ZrZn$_2$, on the
other hand, has not nearly reached its saturation magnetization at ambient
pressure. The above discussion suggests that, as a consequence, the 
superconducting $T_{\rm c}$ in ZrZn$_2$ is likely to monotonically increase
from its onset as one goes deeper into the ferromagnetic phase by decreasing
the pressure, while in UGe$_2$ one expects the superconducting $T_{\rm c}$
to peak close to the magnetic phase boundary, leading effectively only to
a pocket of superconductivity, as shown schematically in Fig.\ \ref{fig:1}(a).
This is indeed what is observed in the experiments, and the present model
provides a qualitative explanation. Consistent with the experimental observation
that the magnetic transition in ZrZn$_2$ is at most very weakly first order,
we have used $t_{\rm min}=0$ in Fig.\ \ref{fig:6}.

\section{DISCUSSION}
\label{sec:V}

We conclude with a summary of our results, and then briefly discuss several
open questions.

\subsection{Summary, and additional remarks}
\label{subsec:V.A}

We have made two contributions with this paper. The first one is general
in nature, and consists in a field-theoretic framework for deriving
strong-coupling equations for superconductivity, regardless of the
mechanism that causes superconductivity. This framework is a natural
extension and generalization of Landau's theory of phase transitions, 
and it has several
advantages, both philosophical and practical ones, over the many-body
diagrammatic formalism that is usually used to describe superconductivity.
One advantage is the fact that the field theory allows for introducing
several order parameters simultaneously in a very natural and transparent
way. In the present case we have used this to obtain a theory of the
interplay between ferromagnetism and spin-triplet, p-wave superconductivity
that deals with both ordering phenomena on equal footing, and within one
unified framework. No input from other theories has been necessary. We
have restricted our description of the superconductivity to the mean-field 
level, but it is obvious how to include fluctuations within this framework. 

In this context we note that for the pairing mechanism we have assumed, or
for any purely electronic mechanism, there is no separation of energy 
scales, and hence no analog of Migdal's theorem. As a result, Eliashberg
theory cannot be expected to produce quantitative results, in constrast
to the case of phonon-induced superconductivity. A rough order-of-magnitude
estimate of $T_{\rm c}$ may nevertheless be of interest. 
Fay and Appel\cite{FayAppel} have given detailed estimates of the
zero-loop $T_{\rm c}$ for ZrZn$_2$. These ranged from 100 $\mu$K to
500 mK, depending on parameter values that are not well known. With a
mid-range value of 5 mK, which is consistent with the observed absence
of superconductivity in the paramagnetic phase, our enhancement mechanism
yields a maximum $T_{\rm c}$ in the ferromagnetic phase of about 500 mK,
which is the same order of magnitude as observed experimentally. 
Regardless of how seriously one is willing to take such estimates, it is
important to note that any corrections due to the inapplicability of 
Migdal's theorem are expected to equally affect both the paramagnetic
and the ferromagnetic phases. They therefore will not spoil the
relative comparison of the two phases that has been our main objective.

We have also restricted
ourselves to discussing the linearized gap equation, but our formalism is
more general, and includes the feedback of a nonvanishing superconducting
order parameter on the magnetic properties. Since the magnetic properties
are responsible for the superconductivity in the first place, 
this feedback will have to be included self-consistently in a theory of 
the superconducting phase. These points will be pursued in future publications.
We further note that we have assumed the superconducting transition
to be of second order. First order scenarios are possible, see 
Ref.\ \onlinecite{Chubukov}.

Our second contribution is specific, and consists of a
calculation of the phase diagram for p-wave, spin-triplet superconductivity
that is driven by ferromagnetic spin fluctuations and coexists with 
ferromagnetism. The most striking qualitative feature
of the result is the pronounced asymmetry of the superconducting critical
temperature in the ferromagnetic and paramagnetic phases, respectively,
which results from the contribution of the magnons in the ferromagnetic
phase to the pairing mechanism. The results are in qualitative agreement
with the experimental observations in UGe$_2$ and ZrZn$_2$.

\subsection{Outlook}
\label{subsec:V.B}

The most intriguing open questions concern the nature of the 
magnetic-fluctuation induced superconducting state, 
and an understanding of the phase
diagram for all temperatures, magnetizations, and magnetic fields. One 
interesting question is whether there is more than one
superconducting phase as a function of temperature,
magnetization, and external magnetic fields. Since the pairing
mechanism is expected to be electronic in origin, and itself sensitive to
superconductivity, it is easy to imagine additional superconducting states
appearing inside the superconducting phase as the temperature is lowered. In
particular, the Cooper pairs that form in the superconducting state produce an
additional internal magnetic field. If this field acts like an external field
for the electrons, then the fermionic spin waves, i.e., the ferromagnetic 
Goldstone modes, will acquire a mass from this effect, leading to a smaller
fluctuation contribution to the $T_{\rm c}$ mechanism. This suggest another
phase inside the superconducting one, where Cooper pairs anti-aligned to the
magnetization might also exist. In general, we also note that the concepts
of transverse and longitudinal critical magnetic fields needs to be worked
out for these superconducting states.

To understand the precise nature of the superconducting state, another
fundamental problem must be addressed. In the theory presented in Section II,
the only coupling of the internal magnetic field in the ferromagnetic phase 
to the superconducting order parameter is through the 
Zeeman-like term that leads to the Stoner splitting of the Fermi surface,
and through the Goldstone modes (which in some sense are also
due to the Zeeman term) that exist in the ferromagnetic
phase. An obvious question is then whether there are orbital effects
in the Cooper channel due to the internal magnetic field.
If there are such effects, they
will lead to an inhomogeneous superconducting state, viz., to some type
of Abrikosov flux-lattice state. This magnetic field effect is of
relativistic origin and has been neglected in the current theory.
It has not yet been discussed theoretically
for pairing mechanisms that are of electronic origin and sensitive to
internal magnetic field effects. 

Finally, our theory of the superconducting phase transition is a mean-field
treatment. The validity of this type of approximation in estimating a
superconducting critical temperature is not obvious. We note, however, that
one would expect any fluctuation-induced degradation of the superconducting 
$T_{\rm c}$ to be larger in the paramagnetic phase than in the ferromagnetic 
one, since for triplet superconductivity, the direction
of the internal magnetic field caused by the magnetization will lead to an
additional phase coherence of the triplet Cooper pairs in the magnetic
phase. The relative importance of this effect on the phase diagram
needs to be investigated.

\acknowledgments
It is a pleasure to thank the Aspen Center for Physics, where this work was
initiated, for its hospitality, as well as our collaborators on an earlier
version of this theory, Thomas Vojta and Rajesh Narayanan.
This work was supported by the NSF under Grant Nos. DMR-98-70597, DMR-99-75259,
DMR-01-32555, and DMR-01-32726.

\appendix
\section{Conventional Eliashberg Theory}
\label{app:A}

In this appendix we show how to derive conventional Eliashberg theory by
means of field-theoretic methods. This is an extension to strong-coupling
theory of previous field-theoretic derivations of BCS 
theory.\cite{PopovNagaosa}

We start with the fermionic field theory defined in Sec.\ \ref{subsec:II.A}, 
but add harmonic phonons,
\bml
\label{eqs:A.1}
\be
S_{\rm ph} = -\frac{1}{2} \int dx\,dy\ \varphi(x)\,D^{-1}(x-y)\,\varphi(y)\quad,
\label{eq:A.1a}
\ee
and an electron-phonon interaction,
\be
S_{\rm e-ph} = g\int dx\ \varphi(x)\sum_{\sigma}{\bar\psi}_{\sigma}(x)\,
               \psi_{\sigma}(x)\quad.
\label{eq:A.1b}
\ee
\eml%
Here $D^{-1}$ is the phonon vertex function whose inverse, $D(x-y) = D(y-x)$, 
is the phonon propagator, and $g$ is the electron-phonon coupling
constant. For simplicity, we consider a scalar phonon field $\varphi(x)$; the
phonon polarizations can be restored in an obvious way if desired.

We now proceed to integrate out the phonons. This
produces an effective electron-electron interaction, which we decompose
into particle-hole and particle-particle channels, and spin-singlet and
spin-triplet contributions in each channel, as usual.\cite{AGD} 
In the particle-particle
channel, only the spin-singlet is nonzero, and in the particle-hole channel
we neglect the spin-triplet contribution $S_{\rm t}^{\rm p-h}$, 
Eq.\ (\ref{eq:2.2e}). The electronic part of the action then reads
\bml
\label{eqs:A.2}
\bea
S &=& -\int dx \sum_{\sigma}{\bar\psi}(x)\,\partial_{\tau}\,\psi(x)
    + S_{0} + S_{\rm s}^{\rm p-h}
\nonumber\\
&&\hskip 30pt + S_{\rm s,e-ph}^{\rm p-h} + S_{\rm s,e-ph}^{\rm p-p}\quad,
\label{eq:A.2a}
\eea
with
\bea
S_{\rm s,e-ph}^{\rm p-p} &=& g^2 \int dx\,dy\ {\bar\psi}_{\downarrow}(x)\,
   {\bar\psi}_{\uparrow}(y)\,D(x-y)\,\psi_{\uparrow}(y)\,
   \psi_{\downarrow}(x)\ ,
\nonumber\\
\label{eq:A.2b}\\
S_{\rm s,e-ph}^{\rm p-h} &=& \frac{-g^2}{4}\int dx\,dy\sum_{\sigma,\sigma'}
   {\bar\psi}_{\sigma}(x)\,\psi_{\sigma}(y)\,D(x-y)\,
\nonumber\\
&&\hskip 50pt \times {\bar\psi}_{\sigma'}(y)\,\psi_{\sigma'}(x)\quad,
\label{eq:A.2c}
\eea
\eml%
and $S_0$ and $S_{\rm s}^{\rm p-h}$ from Sec.\ \ref{subsec:II.A}.

Now we introduce the Nambu spinors
\bml
\label{eqs:A.3}
\be
\Psi(x) = \left(\begin{array}{c} \psi_{\uparrow }(x) \\
                                {\bar\psi}_{\downarrow}(x)
               \end{array}\right)\quad,\quad
{\bar\Psi}(x) = \left({\bar\psi}_{\uparrow}(x),\psi_{\downarrow}(x)\right)
                  \quad,  
\label{eq:A.3a}
\ee
and the corresponding composite variables
\be
\Phi_{ij}(x,y) = {\bar\Psi}_i(x)\,\Psi_j(y)\quad.  
\label{eq:A.3b}
\ee
\eml%

The field $\Phi$ is bilinear in the fermion fields, and hence its components
commute with each other as well as with all other objects. $\Phi$ is therefore
isomorphic to a classical field ${\cal G}$, and we can transform to
a description in terms of these variables by exactly rewriting the
partition function, Eq.\ (\ref{eq:2.1}), as
\bea
Z &=&\int D[{\bar\psi},\psi]\ e^{S[{\bar\psi},\psi]}
      \int D[{\cal G}]\ \delta [{\cal G} - \Phi] 
\nonumber\\
&=&\int D[{\bar\Psi},\Psi]\ e^{S[{\bar\Psi},\Psi]}
    \int D[{\cal G},\Lambda]\ e^{\Tr[\Lambda ({\cal G} - \Phi)]}
\nonumber \\
&=&\int D[{\cal G},\Lambda]\ e^{{\cal A}[{\cal G},\Lambda]}\quad.  
\label{eq:A.4}
\end{eqnarray}
Here $\Lambda$ is an auxiliary field that serves as a Lagrange
multiplier, and we have defined an effective action
\bml
\label{eqs:A.5}
\bea
{\cal A}[{\cal G},\Lambda] &=& 
   \frac{1}{2}\Tr\ln ({\tilde G}_0^{-1} - \Lambda^{\rm T}) 
       + \Tr(\Lambda{\cal G})
\nonumber\\
&&+ g^2\int dx\,dy\ {\cal G}_{12}(x,y)\,D(x-y)\,{\cal G}_{21}(y,x)
\nonumber\\
&&-\frac{g^2}{4}\int dx\,dy\ \bigl[{\cal G}_{11}(x,y)\,D(x-y)\,
                                                       {\cal G}_{11}(y,x)
\nonumber\\
&&\hskip 40pt   + {\cal G}_{22}(x,y)\,D(x-y)\,{\cal G}_{22}(y,x)
\nonumber\\
&&\hskip 40pt   - 2{\cal G}_{11}(x,y)\,D(x-y)\,{\cal G}_{22}(y,x) \bigr]
\nonumber\\
&&+ \frac{\Gamma_s}{2}\int dx\ \left(\tr{\tilde\sigma}_3\,{\cal G}(x,x)
                   \right)^2\quad.
\label{eq:A.5a}
\end{eqnarray}
$\Tr$ denotes a trace over both discrete and continuous labels, 
while $\tr$ traces over discrete labels only.
${\tilde G}_0^{-1}$ is an inverse free-particle Green operator,
\be
{\tilde G}_0^{-1} = -\partial_{\tau} + \sigma_3\left(\frac{\nabla^2}{2m_{\rm e}} 
                    + \mu\right)\quad,
\label{eq:(A.5b)}
\ee
with $\sigma_3$ the third Pauli matrix. The matrix elements of ${\tilde G}_0$
are
\be
{\tilde G}_0(x-y) = \left(\matrix{G_0(x-y) & 0 \cr
                                  0 & -G_0(y-x)}\right)\quad,
\label{eq:A.5c}
\ee
with
\bea
G_0(x-y) &=& \langle{\bar\psi}_{\sigma}(x)\,\psi_{\sigma}(y)\rangle_{S_0}
\nonumber\\
         &=& \langle x\vert-\partial_{\tau} + \nabla^2/2m_{\rm e} 
               + \mu \vert y\rangle\quad,
\label{eq:A.5d}
\eea
\eml%
the usual free particle Green function.

The standard strong-coupling, or Eliashberg, theory of superconductivity is
just the saddle-point solution of the field theory given by 
Eqs.\ (\ref{eqs:A.5}). The saddle-point condition is
\be
\frac{\delta A}{\delta{\cal G}} = 0\quad,\quad
     \frac{\delta A}{\delta\Lambda} = 0\quad.  
\label{eq:A.6}
\ee
We are looking for homogeneous saddle-point solutions
${\cal G}_{\rm sp}(x,y) = {\cal G}(x-y)$ and 
$\Lambda_{\rm sp} (x,y) = \Lambda (x-y)$, and it is easy to see
that the diagonal components of both ${\cal G}$ and $\Lambda$
are related. To express this fact, and to conform with standard
notation,\cite{AGD} we write
\bml
\label{eqs:A.7}
\bea
{\cal G}(x-y) &=& \left(\matrix{G(x-y) & -F^+(x-y) \cr
                               -F(x-y) & -G(y-x) \cr}\right)\quad,
\label{eq:A.7a}\\
\Lambda(x-y) &=& \left(\matrix{\Sigma(x-y) & \Delta^+(x-y) \cr
                               \Delta(x-y) & -\Sigma(y-x) \cr}\right)\quad.
\label{eq:A.7b}
\eea
\eml%
As we will see, $G$ and $F$, $F^+$ are the usual normal and anomalous Green
functions, and $\Sigma$ and $\Delta$, $\Delta^+$ are the normal and anomalous
self energies, respectively.

The first equality in Eq.\ (\ref{eq:A.6}) gives
\bml
\label{eqs:A.8}
\bea
\Sigma(x-y)&=&\left[g^2\,D(x-y) - 2\Gamma_s\,\delta (x-y)\right]\,G(x-y)\quad,
\nonumber\\  
\label{eq:A.8a}\\
\Delta^+(x-y) &=& g^2\,D(x-y)\,F^+(x-y)\quad,  
\label{eq:A.8b}\\
\Delta(x-y) &=& g^2\,D(x-y)\,F(x-y)\quad.
\label{eq:A.8c}
\eea
\eml%
The second equality yields
\bml
\label{eqs:A.9}
\bea
G(x-y) &=& G_0(x-y) 
\nonumber\\
&&\hskip -20pt + \int dx'\,dy'\ G_0(x-x')\,\Sigma(x'-y')\,G(y'-y)
\nonumber\\
&&\hskip -20pt + \int dx'\,dy'\ G_0(x-x')\,\Delta(x'-y')\,F^+(y'-y) \ ,
\nonumber\\
\label{eq:A.9a}\\
F^+(x-y) &=& \int dx'\,dy'\ G_0(x'-x)\,\Sigma(y'-x')\,F^+(y'-y)
\nonumber\\
&&\hskip -20pt + \int dx'\,dy'\ G_0(x'-x)\,\Delta^+(x'-y')\,G(y'-y) \ ,
\nonumber\\
\label{eq:A.9b}\\
F(x-y) &=& \int dx'\,dy'\ G_0(x-x')\,\Sigma(x'-y')\,F(y'-y)
\nonumber\\
&&\hskip -20pt + \int dx'\,dy'\ G_0(x-x')\,\Delta(x'-y')\,G(y-y')\ .
\nonumber\\
\label{eq:A.9c}
\eea
\eml%
The Eqs.\ (\ref{eqs:A.8}) and (\ref{eqs:A.9}) are the standard 
Eliashberg equations for conventional, spin-singlet, 
phonon-induced superconductivity.\cite{AGD} 
$\Gamma_s$ plays the role of the Coulomb pseudopotential.\cite{AndersonMorel}

The linearized gap equation, which determines the superconducting transition
temperature, is obtained by expanding the Eliashberg equations to linear order
in $\Delta$. In Fourier space, with frequency-momentum four-vectors
$k=(i\omega_n,{\bf k})$, one finds for the linearized gap equation
\bml
\label{eqs:A.10}
\be
\Delta(k) = g^2\int_q D(k-q)\,G(q)\,\Delta(q)\,G(-q)\quad,
\label{eq:A.10a}
\ee
with a normal Green function
\be
G(k) = 1/\left[G_0^{-1}(k) -\Sigma(k)\right]\quad,
\label{eq:A.10b}
\ee
in terms of a normal self energy
\be
\Sigma(k) = g^2\int_q D(k-q)\,G(q) - \Gamma_{\rm s}n_{\rm e}\quad,
\label{eq:A.10c}
\ee
\eml%
Here $\int_q \equiv T\sum_n\int d{\bf q}/(2\pi)^2$,
and $n_{\rm e} = 2\Tr G$ is the electron number density.

\end{document}